\documentclass[useAMS]{mn2e}
\usepackage{psfig,epsfig}
\usepackage{graphicx}
\usepackage{amssymb}
\usepackage{epstopdf}

\begin{document}
\def\simlt{\mathrel{\rlap{\lower 3pt\hbox{$\sim$}}
        \raise 2.0pt\hbox{$<$}}}
\def\simgt{\mathrel{\rlap{\lower 3pt\hbox{$\sim$}}
        \raise 2.0pt\hbox{$>$}}}

\title[Infrared properties of radio AGN]{The PEP Survey: Infrared Properties of Radio-Selected AGN}

\author[Manuela Magliocchetti et al.]
{\parbox[t]\textwidth{M. Magliocchetti$^{1}$,  D. Lutz$^{2}$, D. Rosario$^{2}$, S. Berta$^{2}$, E. Le Floc'h$^{3}$, B. Magnelli$^{2}$,   F. Pozzi$^{4}$, L. Riguccini$^{3}$, P. Santini$^{5}$\\
} \\
{\tt $^1$ INAF-IAPS, Via Fosso del Cavaliere 100, 00133 Roma,
  Italy}\\
  {\tt $^2$ Max Planck Institute f\"ur Extraterrestrische Physik (MPE),
  Postfach 1312,  D85741, Garching, Germany}\\
  {\tt $^3$ CEA-Saclay, Service d'Astrophysique, F-91191,
  Gif-sur-Yvette, France}\\
  {\tt $^4$ Dipartimento di Astronomia, Universita' di Bologna, Via Ranzani 1, 40127, Bologna, Italy}\\
  {\tt $^5$ INAF-Osservatorio Astronomico di Roma, Via di Frascati 33, 00040 Monte Porzio Catone, Italy}\\
%{\tt $^2$ Dip. Fisica, Universita' 'Tor Vergata', Via Ricerca Scientifica 1, 00133, Roma Italy}\\
%{\tt $^3$ SISSA-ISAS, Via Bonomea, 265, 34136 Trieste, Italy}\\
%{\tt $^4$ INAF-Osservatorio Astronomico di Padova, Vicolo Osservatorio 5, 35122 Padova, Italy} 
}
\maketitle
\begin{abstract}
By exploiting the VLA-COSMOS and the {\it Herschel}-PEP surveys, we investigate the Far Infrared (FIR) properties of radio-selected AGN. To this purpose, from VLA-COSMOS we considered the 1537,  $F_{1.4 \rm GHz}\ge0.06$  mJy sources  with a reliable redshift estimate, and sub-divided them into star-forming galaxies and AGN {\it solely} on the basis of their radio luminosity. 
The AGN sample is {\it complete} with respect to radio selection at all $z\simlt 3.5$.
832 radio sources have a counterpart in the PEP catalogue. 175 are AGN. Their redshift distribution closely resembles that of the total radio-selected AGN population, and exhibits two marked peaks at $z\sim 0.9$ and $z\sim 2.5$.
We find that the probability for a radio-selected AGN to be detected at FIR wavelengths is both a function of radio power and redshift, whereby powerful sources are more likely to be FIR emitters at earlier epochs. 
This is due to two distinct effects: 1) at all radio luminosities, FIR activity monotonically increases with look-back time and 2)  radio activity of AGN origin is increasingly less effective at inhibiting FIR emission. 
Radio-selected AGN with FIR emission are preferentially  located in galaxies which are smaller than those hosting FIR-inactive sources. Furthermore, at all $z\simlt 2$, there seems to be a preferential (stellar) mass scale M$_*\sim [10^{10}-10^{11}] \rm M_\odot$  which maximizes the chances for FIR emission.
We find such FIR (and MIR) emission  to be due to processes  indistinguishable from those which power star-forming galaxies.  It follows that radio emission  in at least 35\% of the entire AGN population is  the sum of two contributions: AGN accretion and  star-forming processes within the host galaxy. 
\end{abstract}

\section{Introduction}
Radio sources constitute an unequally powerful tool to investigate the origin and evolution of our Universe. In fact, radio emission is not obscured or attenuated by dust or gas, so deep radio surveys offer a unique opportunity to detect and study sources up to the highest redshifts.   It is following this rationale that very ambitious programs like the planned Square Kilometer Array (SKA, Carilli et al., 2004) facility, a gigantic radio telescope with a total collective area of 1 square kilometer, or its precursors ASKAP (australian SKA pathfinder, Johnston S. et al. 2007) and MeerKAT (Jonas J.L. 2009) will soon see their first light. When applied to radio sources, the expected advantages in sensitivity, field-of-view, frequency range and spectral resolution will yield terrific progress in many research fields, from cosmology to astrophysics.

Extra-galactic radio sources are however a mixed bag of astrophysical objects:  amongst the most powerful ones we find radio-loud QSOs and FRII (Fanaroff \& Riley 1974) galaxies while, moving to weaker sources, the dominant populations become FRI,  low-excitation galaxies and  star-forming galaxies, whose contribution to the total radio counts become predominant at the sub-mJy level.
Discerning amongst these objects and their relative weight is an impossible task to carry out only on the basis of radio-continuum data. This is why increasingly more effort was recently put on photometric and spectroscopic follow-up studies of radio sources detected in deep-field surveys (amongst the many: Schinnerer et a. 2004; 2007; 2010 for the COSMOS field;  Morrison et al. 2010 for GOODS-N; Mc Alpine et a. 2013 for the VIDEO-XMM3 field;  Bondi et al. 2003 for the VVDS field: Simpson et al. 2012 for the Subaru/XMM-Newton Deep Field: Mao et al. 2012 for the ATLAS field). 
Investigations of the optical properties of radio sources can then disentangle the different populations and also allow to determine a number of quantities which influence the source behavior like e.g. the black-hole mass in optical spectra dominated by broad emission lines (e.g. Metcalf \& Magliocchetti 2006).   

Following the launch of the {\it Spitzer} satellite, in the recent years radio sources have also been investigated at Mid-Infrared (MIR) wavelengths (e.g. Appleton et al. 2004; Boyle et al. 2007; Magliocchetti, Andreani \& Zwaan 2008; Garn et al. 2009;  Leipski et al. 2009; De Breuck et al. 2010; Norris et al. 2011 amongst the many). These studies proved to be extremely useful for two main reasons: they allowed  investigations of the properties and evolution of the sub-population of star-forming galaxies up to $z\sim 2$, while on the other hand they  provided interesting hints on the  central engine responsible for radio (and MIR) AGN emission.

It is only in these very last few years that the radio community has started to realize the importance of multi-wavelength studies which also include Far-Infrared (FIR) information. This growing interest is primarily due to the advent and launch of the {\it Herschel} satellite, which for the first time has observed galaxies at FIR wavelengths up to very large cosmological ($z\sim 4$) distances. Earlier FIR missions such as IRAS, ISO and {\it Spitzer}@70$\mu$m only probed the {\bf relatively local ($z\simlt 1$)} universe, and since AGN-powered radio galaxies in the local universe are almost ubiquitously hosted by massive red galaxies with little or no ongoing star-formation activity (e.g. Magliocchetti et al. 2002; 2004), and since FIR emission in galaxies is almost entirely due to processes connected with ongoing star formation, it follows that any pre-{\it Herschel}  interest in a joint radio-FIR analysis was limited to re-assess the precision of the tight radio-FIR correlation found by early IRAS-based studies of local star-forming galaxies (Condon et al. 1982).

However, when moving to higher redshifts, galaxies undergo dramatic changes.  For instance, it is now well assessed that both the cosmic star-forming activity and AGN output steadily increase with look-back time at least up to redshifts $z\sim2$ (e.g. Gruppioni et al. 2013: Merloni, Rudnick \& Di Matteo 2004), and many galaxies are observed to host both an active AGN and ongoing star-formation (e.g. Alexander et al. 2008). It is then legitimate to wonder whether radio galaxies also follow such a trend and if in the early universe they are also associated with ongoing stellar production. 
First attempts in this direction were presented by the work of Seymour et al. (2011) which  analyse the FIR emission as observed by the SPIRE instrument (Griffin et al. 2010) on board of the {\it Herschel} satellite (Pilbratt et al. 2010) for a sample of very powerful, 
L$_{1.4 \rm GHz}\ge 10^{25}$ W Hz$^{-1}$ radio sources selected in the extra-galactic {\it Spitzer} First Look Survey field, and by the work of Del Moro et al. (2013) who investigate hidden AGN activity in a sample of GOODS-North sources with deep radio, infrared and X-ray data.

The aim of this paper is to push the Seymour et al. (2011) results further and investigate, for the first time,  the FIR properties of radio-selected AGN of {\it all} radio luminosities and at {\it all} redshifts $z\simlt 3.5$. This will be done by taking advantage of the very deep radio catalogue obtained at 1.4 GHz in the COSMOS field by Schinnerer et al. (2004; 2007) and Bondi et al. (2008). AGN will be selected solely on the basis of their radio luminosity by following the results of Mc Alpine et al. (2013). FIR information for such sources will be provided by the PEP survey (Lutz et al. 2011), proven to be deep enough to provide extremely important results on the ongoing processes within the AGN hosts and on their cosmological evolution.
Throughout this paper we assume a $\Lambda$CDM cosmology with $H_0=70 \: \rm km\:s^{-1}\: Mpc^{-1}$ ($h=0.7$), $\Omega_0=0.3$,  $\Omega_\Lambda=0.7$ and $\sigma^m_8=0.8$.

%%%%%%%%%%%%%%%%%%%%%%%%%%%%%%%%%%%%%%%%%%%%%%%%%%%%%%%%%%%%%%%%%%%%%%%%%%%%%%%%%%%%%%%%%%%%%%%%%%%%%%%%%%%%%%%%%%%%%%%%%%%%%%

\section{the Radio-Infrared Master catalogue}
\subsection{Radio data}
The VLA-COSMOS Large Project observed the 2 deg$^2$ of the  COSMOS field at 1.4 GHz with the VLA in the A configuration. This survey,  extensively described in Schinnerer et al. (2004) and Schinnerer et al. (2007), provides continuum radio observations with a resolution of 2$^{\prime\prime}$ and a mean 1$\sigma$ sensitivity of about 10.5 $\mu$Jy in the innermost 1 deg$^2$ region and of about 15 $\mu$Jy in the outer parts.

The catalogue adopted in our work is that presented in Bondi et al. (2008), which comprises 2417 sources selected above the 5$\sigma$  threshold. This roughly corresponds to a 1.4GHz integrated flux of 60 $\mu$Jy. 
From the catalogue, we removed those sources flagged as potential sidelobes. This brings the number of radio objects considered for our analysis down to 2382.

78 of the above sources clearly present a complex structure, owing to the presence of jets and lobes. Of these, 60 out of 78 have integrated fluxes brighter than 1 mJy.
The Schinnerer et al. (2007) and Bondi et al. (2008) catalogues flag these sources and list their centers as either that of the radio core or of the optical counterpart when either of these could be identified, or as the luminosity weighted position in all other cases..

In order to provide radio sources with a redshift determination, we cross correlated the above sample with the Ilbert et al. (2013) catalogue of redshifts (either photometric or, whenever possible, spectroscopic) provided for COSMOS galaxies. Given the high positional accuracy of both the radio and the optical-near infrared surveys, we fix the matching radius to 1 arcsec. This procedure provides redshift estimates for 1537 radio sources, i.e.  $\sim 65$ per cent of the parent sample, with  only a very small fraction (about 3.8\%) expected to be spurious matches. Note that the above percentage is roughly independent of radio flux. In fact, if we consider catalogues including sources respectively brighter than 0.1, 0.25, 0.5 and 1 mJy, we obtain redshift estimates for 64\%, 64\%, 61\% and 58\% of the parent samples. These values are summarized in the second and third columns of Table 1. Their visual representation is provided in Figure \ref{fig:ids} by the empty circles, where the errors associated with the points are Poissonian estimates.

\subsection{Infrared data}

\begin{figure}
\begin{center}
%\vspace{8cm}  % amount of vertical space needed
\includegraphics[scale=0.45]{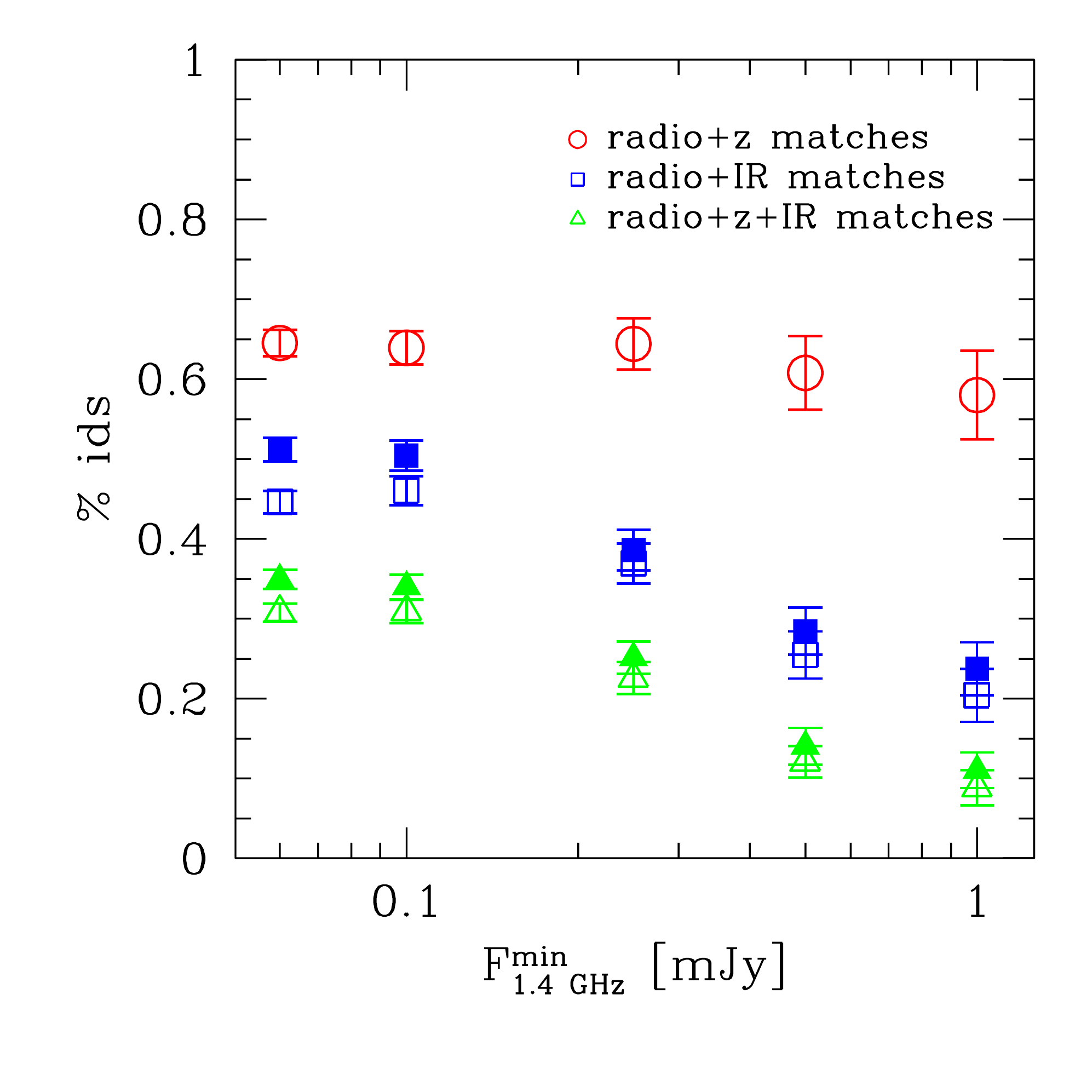}
\caption{Fraction of VLA-COSMOS radio sources with an estimated redshift (red, empty circles), with a counterpart in the PEP catalogue (blue squares) and with both a PEP counterpart and an estimated redshift (green triangles) as a function of minimum 1.4 GHz flux. Empty triangles and squares indicate sources with a counterpart at 100$\mu$m, while filled ones are for radio objects with a counterpart either at 100$\mu$m or at 160$\mu$m. Errorbars correspond to Poissonian estimates.
\label{fig:ids}}
\end{center}
\end{figure}

\begin{table}
\begin{center}
\caption{Properties of the analysed sample. The various columns represent: 1) radio flux cut in mJy units; 2) total number of COSMOS-VLA sources with fluxes brighter than the chosen flux limit, N$_{\rm TOT}$; 3) number of radio sources with a redshift estimate, N$_{\rm z}$; 4) number of radio sources with a counterpart at 100$\mu$m from the PEP catalogue, N$_{100{\rm \mu m}}$; 5) number of radio sources with a counterpart in the PEP catalogue either at 100$\mu$m or at $160 \mu$m, N$_{\rm IR}$; 6) number of radio sources with both an infrared counterpart (either at 100$\mu$m or at 160 $\mu$m) and a redshift estimate,  N$_{\rm zIR}$. A visual representation of the Table is provided in Figure \ref{fig:ids}. }
\begin{tabular}{llllll}
Flux cut [mJy]  & N$_{\rm TOT}$&N$_{\rm z}$  & N$_{100{\rm \mu m}}$& N$_{\rm IR}$ &N$_{\rm zIR}$\\
\hline
\hline
1& 181& 105& 37&43&20\\
0.5& 306& 186&78&87&43\\
0.25& 593& 382&219&239&149\\
0.1& 1436& 918&661&724&488\\
0.06&2382&1537&1063&1219&832\\
\hline
\hline

\end{tabular}
\end{center}
\end{table}

\begin{figure}
\begin{center}
%\vspace{8cm}  % amount of vertical space needed
\includegraphics[scale=0.45]{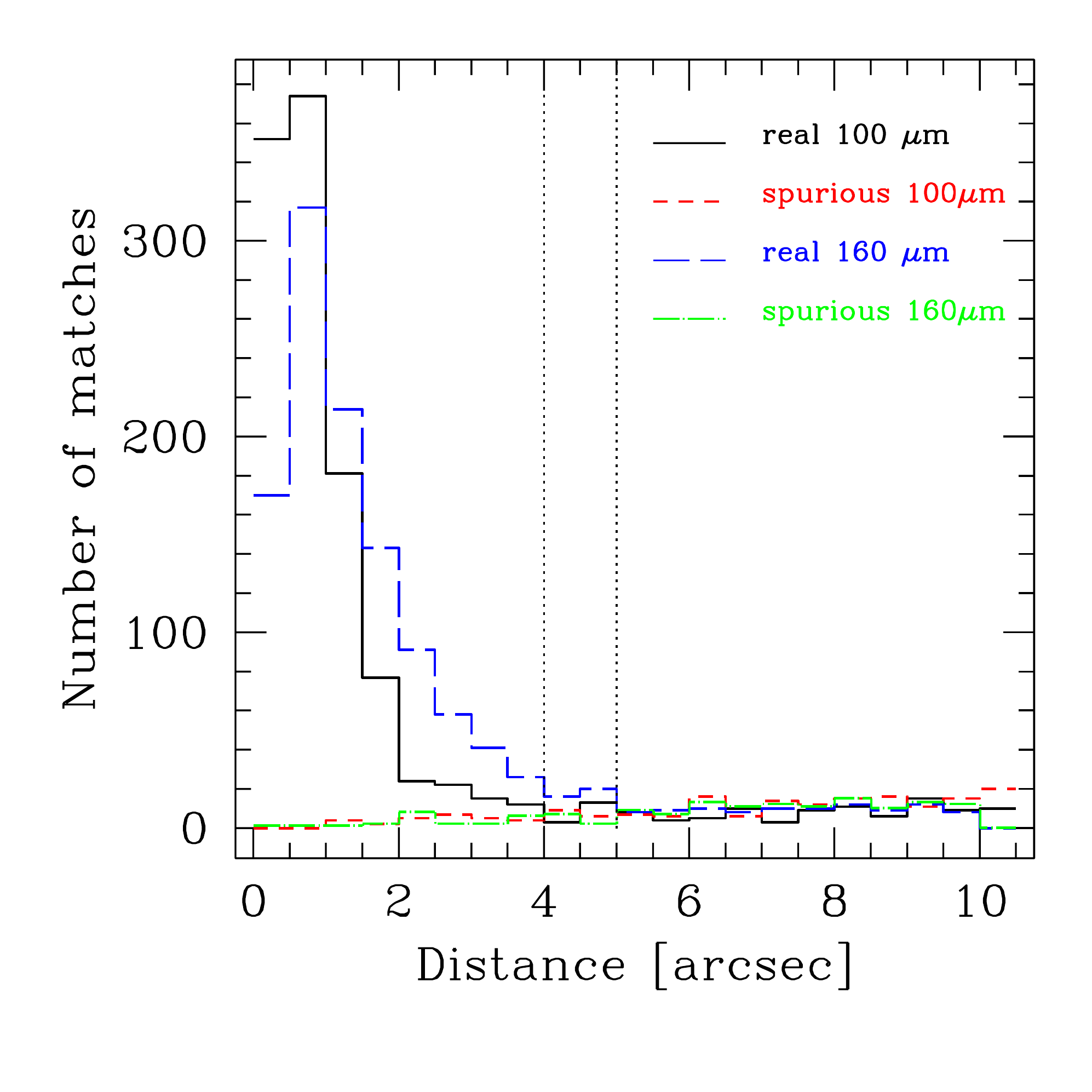}
\caption{Number of PEP counterparts to F$_{1.4 \rm GHz}\ge 60 \mu$Jy VLA-COSMOS radio objects as a function of source separation. Spurious matches have been estimated by vertically shifting all the infrared positions by 1 arcmin. The two vertical dotted lines represent the maximum radii (respectively 4$^{\prime\prime}$ at 100$\mu$m and 5$^{\prime\prime}$ at 160$\mu$m) allowed for the matching procedure.
\label{fig:match}}
\end{center}
\end{figure}

The COSMOS region has been observed by the PACS (Poglitsch et al. 2010) instrument onboard the {\it Herschel} Space Observatory (Pilbratt et al. 2010) as a part of the PACS Evolutionary Probe (PEP,  D. Lutz et al. 2011) Survey, aimed at  studying the properties and cosmological evolution of the infrared population up to redshifts $z\sim 3-4$. 
 We refer to the Lutz et al. (2011) and Berta et al. (2010)  papers for further information on the survey and fields.

The total number of infrared sources detected in COSMOS in the blind (i.e. without the use of priors, in order to minimize the possible bias effects introduced by a double selection) catalogues at the  $\simgt 3\sigma$ confidence level is  5355 at 100$\mu$m and 5105 at 160$\mu$m.  The corresponding flux limits are $\sim 4$ mJy at 100$\mu$m and $\sim 7$ mJy at 160$\mu$m.  

Infrared counterparts to F$_{1.4 \rm GHz}\ge 60\mu$Jy VLA-COSMOS sources have been found by a simple matching technique between the radio and the COSMOS-PEP catalogues. 
The relatively poor angular resolution of the {\it Herschel} mirror at the two considered wavelengths (respectively $\sim 8^{\prime\prime}$ at 100$\mu$m and $\sim 13^{\prime\prime}$ at 160$\mu$m) introduces some issues regarding  the choice of the optimal maximum radius to be adopted in the matching procedure. In fact, matching radii as large as the above angular resolutions would imply a very large fraction of mismatches (i.e. spurious associations). On the other hand, a too small radius would cause a sensible loss of true associations. 

In order to investigate the above issue, we have then considered the distribution of radio-to-infrared matches as a function of angular separation, both for 100$\mu$m- and for 160$\mu$m-selected sources. This is reproduced in Figure \ref{fig:match} by the solid line for associations with 100$\mu$m-selected sources and by the long dashed line for  radio-to-160$\mu$m associations. The two almost flat distributions at the bottom represent the expected number of mismatches, obtained by vertically shifting all the infrared positions by 1 arcmin. As expected, the distributions of real associations have a peak and a declining tail which then tends to flatten out  at the same level of the spurious matches. Investigations of Figure \ref{fig:match} show that this happens at $\sim 4^{\prime\prime}$ for infrared sources selected at 100$\mu$m, and at $\sim 5^{\prime\prime}$ 
in the case of objects selected at 160$\mu$m. These will be the chosen maximum separation values adopted in the process of matching radio and PEP sources: 4$^{\prime\prime}$ at 100$\mu$m and 5$^{\prime\prime}$ at 160$\mu$m. 

By adopting these matching radii, we find that 1063 VLA-COSMOS sources (corresponding to 44\% of the parent sample) have a counterpart at 100$\mu$m and 1100 (corresponding to 46\%) have a counterpart at 160$\mu$m. The above choice ensures a negligible contribution from spurious associations in our matched catalogues: 2.5\% at 100$\mu$m and 2.9\% at 160$\mu$m. 

Most of the considered radio sources have detections at both 100$\mu$m and 160$\mu$m, so the above numbers are somewhat redundant. 
The final, total number of F$_{1.4 \rm GHz}\ge 60 \mu$Jy VLA-COSMOS sources with an infrared counterpart either at 100$\mu$m or at 160$\mu$m is 1219, corresponding to 51\% of the original sample. 
Of these sources, 832 also have a redshift determination obtained as explained in Section 2.1. The above numbers are summarized in Table~1.

It is interesting to note that, while the fraction of radio sources with a redshift determination presents a distribution which is independent of radio flux, the number of infrared counterparts to radio-selected sources dramatically decreases for increasing radio flux cuts. In fact, by considering radio sources  respectively brighter than 0.1 mJy, 0.25 mJy, 0.5 mJy, and 1 mJy, we find that the percentages of true associations are 50\%, 39\%, 28\% and 24\%. These values are all reported in Table~1 and visually represented in Figure \ref{fig:ids}.
The above issue will be investigated at greater length in the next Section.

\section{Sample Selection}

\begin{figure}
\begin{center}
%\vspace{4cm}  % amount of vertical space needed
\includegraphics[scale=0.45]{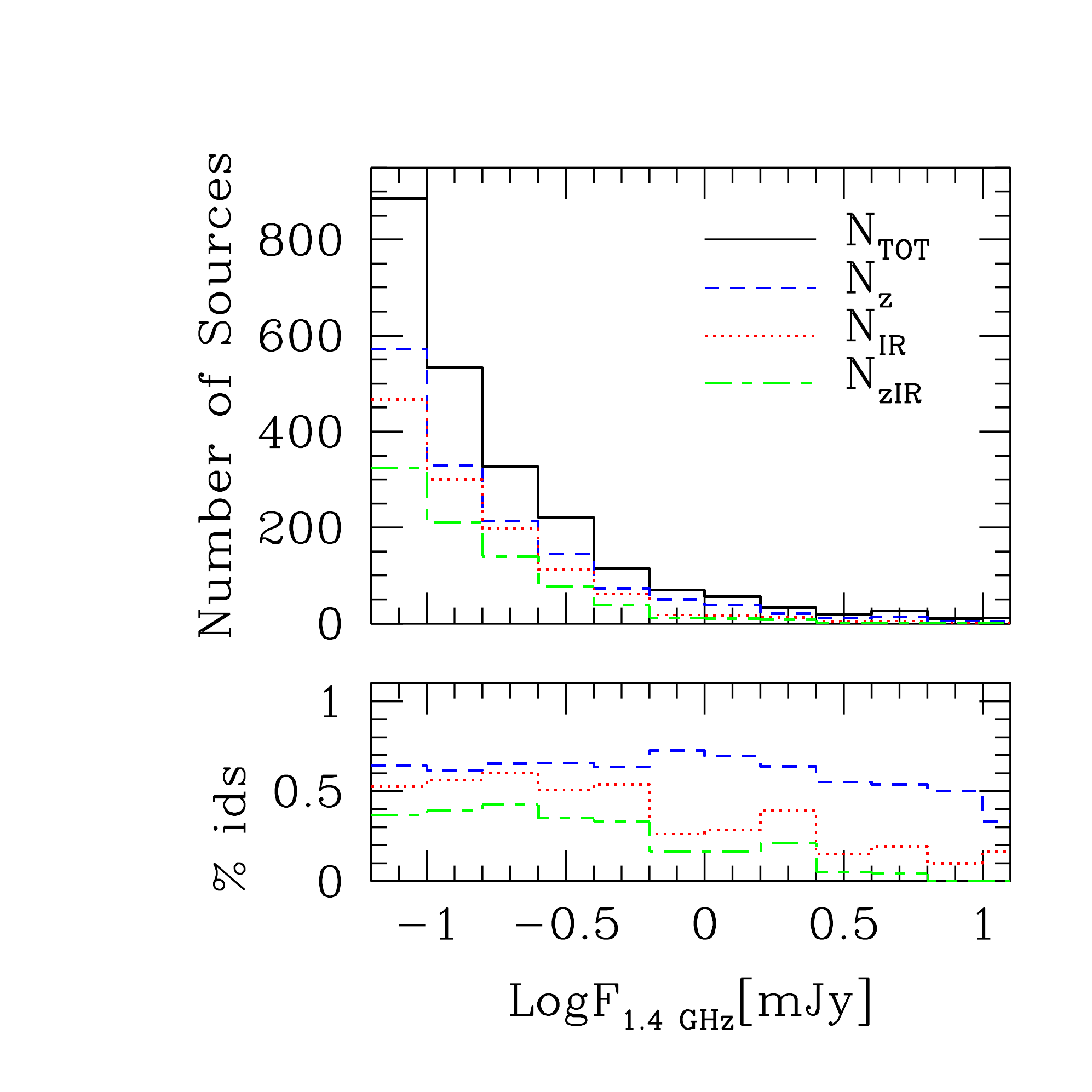}
\caption{Differential number counts of F$_{1.4\rm GHz}\ge 60$ $\mu{\rm Jy}$ radio sources  as a function of radio flux (top panel). The solid line corresponds to all sources in the COSMOS-VLA catalogue, the dashed line to those with a redshift estimate, the dotted line to those with a counterpart in the PEP catalogue either at 100$\mu$m or at $160 \mu$m and the long-short dashed line to those with both a redshift estimate and a PEP id. The bottom panel presents the ratios between these last three quantities and the total number of radio sources in the same flux bin.
\label{fig:idsvss}}
\end{center}
\end{figure}

\begin{figure}
\begin{center}
%\vspace{4cm}  % amount of vertical space needed
\includegraphics[scale=0.45]{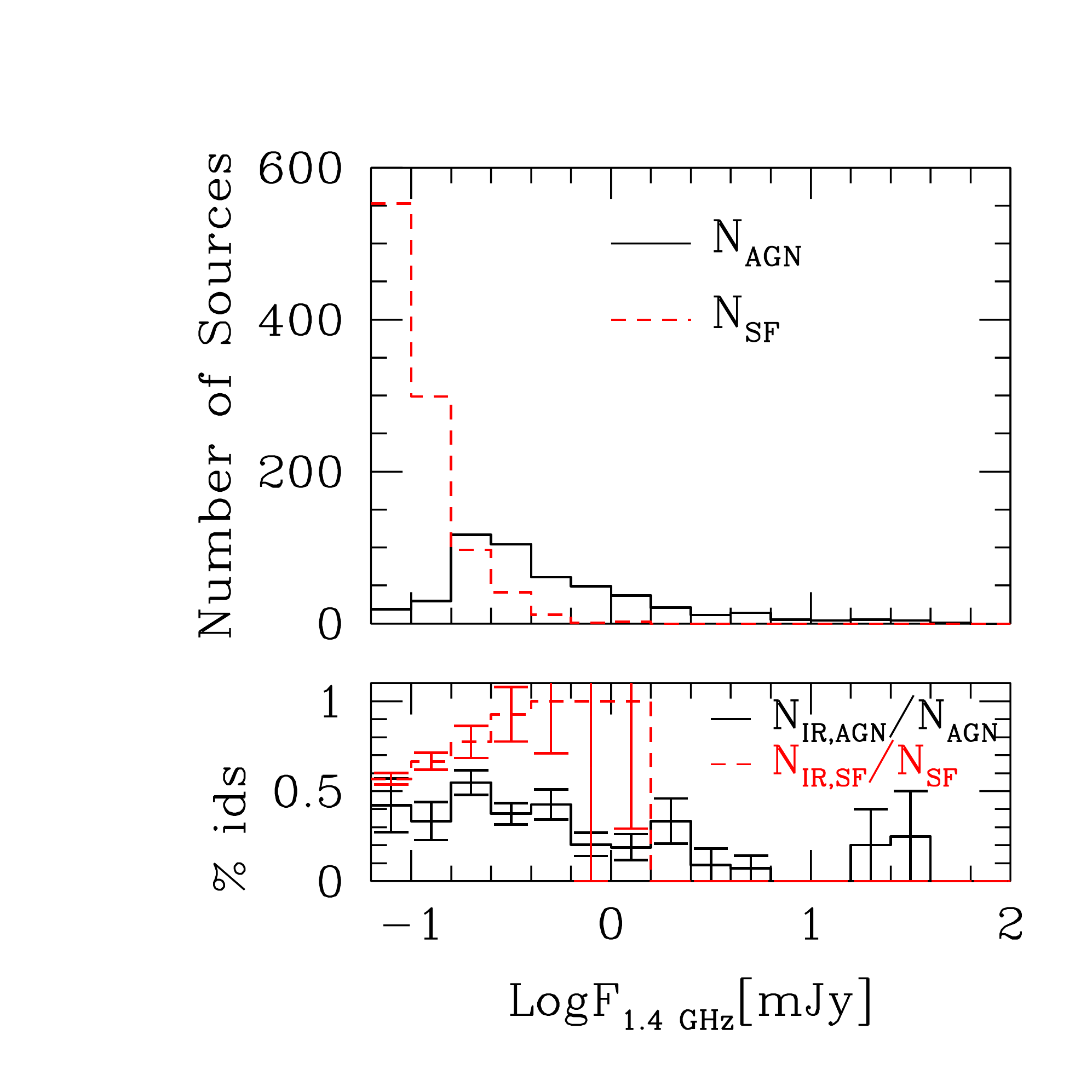}
\caption{Top panel: flux distribution of $\rm F_{\rm 1.4 GHz} \ge 0.06$ mJy COSMOS-VLA sources endowed with a redshift determination. The dashed line represents those objects classified by our method as star-forming galaxies, while the solid line represents AGN. Bottom: percentage of the above sources which also have a FIR counterpart in the PEP catalogues. Again, the dashed line is for star-forming galaxies, while the solid histogram corresponds to AGN. Errorbars correspond to Poissonian estimates.
\label{fig:idsvsstype}}
\end{center}
\end{figure}

A better visualization of the issues raised in \S2 is provided by Figure \ref{fig:idsvss}, which represents the differential number counts of radio sources with and without a redshift estimate and with or without an infrared counterpart.
In fact, Figure \ref{fig:idsvss} (lower panel) clearly shows that while the process of providing a redshift to a radio source is independent of its radio flux (at least down to 60~$\mu$Jy), the same cannot be said for the chances of a radio source to also be a FIR emitter. In this latter case, one has that the probability sharply drops from about 50\% to $\sim$25\% and lower for radio fluxes brighter than $\sim$0.5 mJy. Interestingly enough, we also find that the fraction of radio sources with associated FIR emission is almost constant below the above flux threshold, while it tends to monotonically decrease with radio flux at brighter fluxes. This is probably not surprising. F$_{1.4 \rm GHz}\sim 0.5$ mJy is in fact expected to mark the transition between a regime where radio emission from extra-galactic sources is dominated by AGN accretion processes and that where radio emission is to be mainly attributed to star formation activity (e.g. White et al. 2012 and references therein).

But when in absence of spectroscopic information, how can we distinguish between radio emission of AGN origin and that instead due to star forming processes?
In the local universe, things are quite straightforward. Radio-loud sources are ubiquitously associated to massive elliptical galaxies with little or no  ongoing star formation activity (e.g. Magliocchetti et al. 2000; Magliocchetti et al. 2002). They are "red and dead", present radio-to-optical (R band) ratios (defined as q${\rm _R=F_{1.4 \rm GHz}\cdot 10^{(R-12.5)/2.5}}$, with F expressed in mJy) larger than $\sim 30$ (e.g. Urry \& Padovani 1995) and show a remarkable tightness in the  K-band magnitude vs redshift correlation (i.e., they are standard candles; Lilly \& Longair 1984), at least up to $z\sim 1$ (Simpson et al. 2012). On the other hand, star-forming galaxies present a very tight correlation between their far-IR and radio fluxes (e.g. Condon et al. 1982; Dickey \& Salpeter 1984), supposedly due to the presence of massive stars which on the one hand reheat dust, therefore producing IR emission, while on the other generate supernovae events which accelerate cosmic rays, therefore producing radio synchrotron radiation (see e.g. Condon 1992).

However, at high redshifts,  discerning between these two populations becomes more uncertain. The pioneering work of Appleton et al. (2004)  uses the radio-to-mid-infrared ratio $q_{24}$ to single out radio emission from star-forming sources and separate it from that due to the the presence of  an AGN up to redshifts $z\sim 1.5$. A similarly tight correlation between far-IR and radio fluxes is also found in the most recent works based on {\it Herschel} and {\it Spitzer} data (e.g. Sargent et al.  2010; Ivison et al. 2010, Bourne et al. 2011, Mao et al. 2011).
At redshifts higher than $z\sim 1.5$ though, all the works based on radio and infrared detections notice a larger spread and/or an evolution of the radio-to-IR ratios from the values which are locally measured note however the opposite finding of the Mao et al. 2011 work, although only based on 10 sources with $z\ge 1.5$). The explanation for this effect is rather simple: $z\sim 1.5$ in fact marks the transition from a relatively 'quiet' universe, to one dominated by AGN and cosmic star-formation activities (e.g. Merloni, Rudnik, Di Matteo 2004). In such a universe, many star-forming galaxies are found to host an AGN (e.g.  Alexander et al. 2008; Del Moro et al. 2013), so a clear distinction between these two classes of sources based on their radio-to-infrared colours is extremely difficult. 
At the same time, galaxies at $z\sim 2$ are generally younger than those found in the local universe, so that also an optical search for radio galaxies as red-and-dead sources might prove to be ineffective. 
For the above reasons, we decided to use radio emission as the sole indicator of AGN rather than star forming activity in the VLA-COSMOS sample. 
 
 \begin{figure*}
\begin{center}
%\vspace{4cm}  % amount of vertical space needed
\includegraphics[scale=0.4]{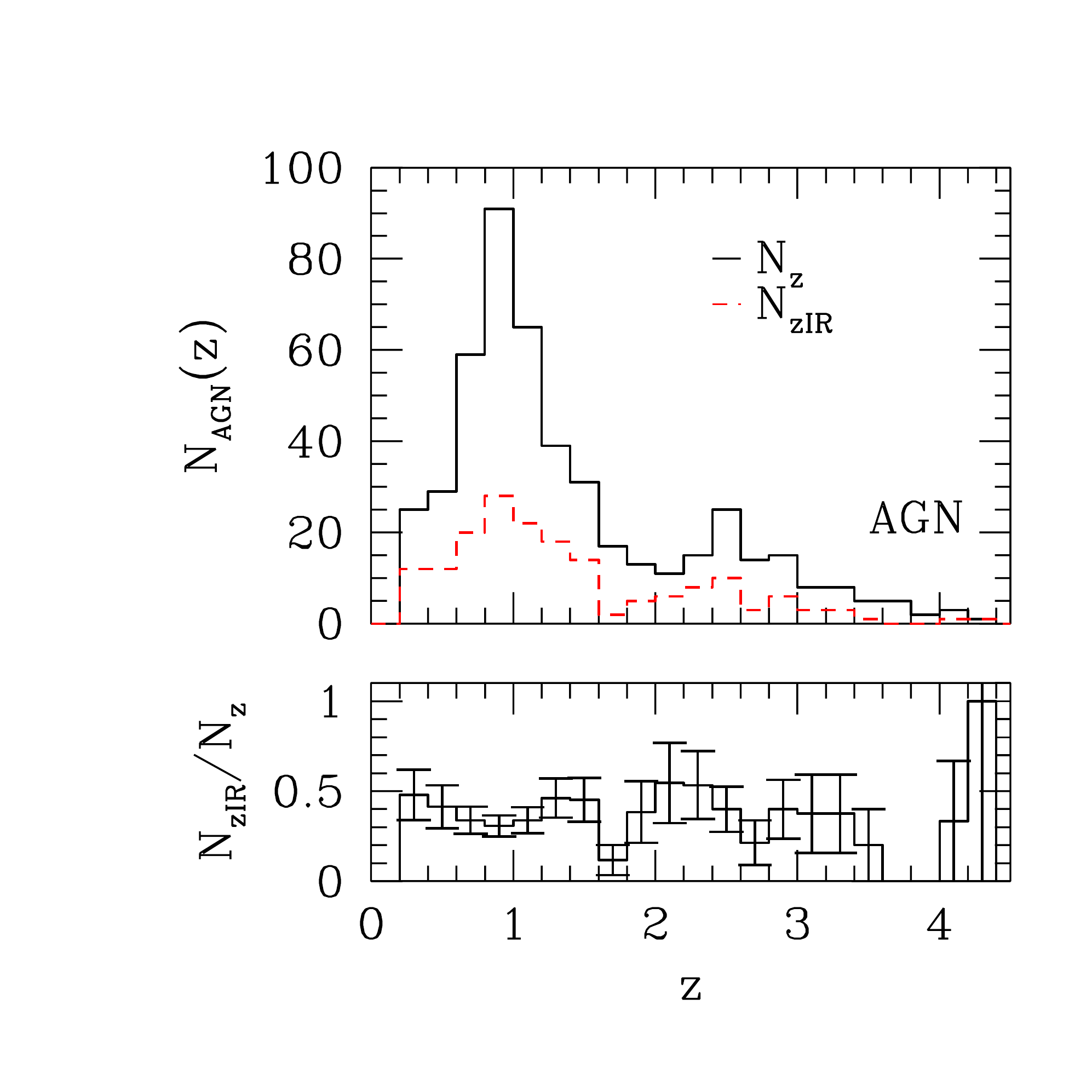}
\includegraphics[scale=0.4]{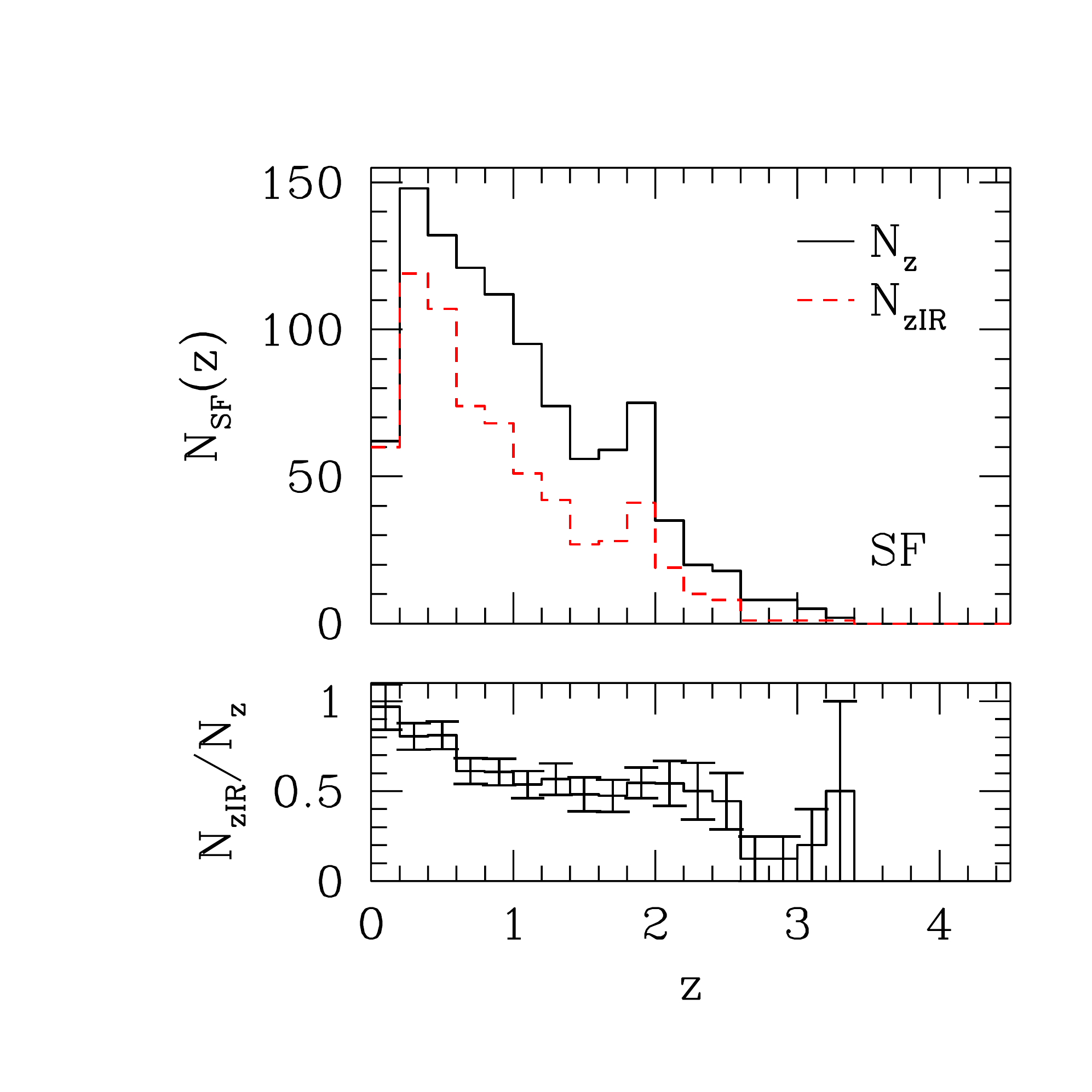}
\caption{COSMOS-VLA redshift distributions for the classes of AGN (left-hand panel) and star-forming galaxies (right-hand panel) brighter than F$_{\rm 1.4 GHz}\ge 0.06$ mJy. The solid lines represent all sources, irrespective of their FIR emission, while the dashed lines indicate those objects which are also FIR emitters. The bottom panels highlight the ratio between these two latter quantities. Errorbars correspond to Poissonian estimates.
\label{fig:idsvsztype}}
\end{center}
\end{figure*}
 
Our approach is based on the  results of McAlpine, Jarvis \& Bonfield (2013) who used the optical and near infrared Spectral Energy Distributions of a sample of 942 radio sources  (out of 1054 objects selected at 1.4 GHz, with a completeness level of 91\%) from the VIDEO-XMM3 field to distinguish between star forming and AGN-powered galaxies and derive their redshifts. 
These authors provide luminosity functions for the two classes of sources up to redshifts $\sim 2.5$ and find different evolutionary behaviours, with star-forming objects evolving in a much stronger way ($\propto (1+z)^{2.5})$ than radio-selected AGN ($\propto (1+z)^{1.2}$). We note that these results are in agreement with those of Smolcic et al. (2009 a,b) obtained on the same COSMOS-VLA dataset we are using in this work.\\
Investigations of their results show that the radio power P$_{\rm cross}$ beyond which  AGN-powered galaxies become the dominant radio population scales with redshift roughly as
\begin{eqnarray}
\rm Log_{10}P_{\rm cross}(z)=\rm Log_{10}P_{0,\rm cross}+z,
\label{eq:P}
\end{eqnarray}
at least up to $z\sim 1.8$. $P_{0,\rm cross}=10^{21.7}$[W Hz sr$^{-1}$] is the value which is found in the local universe and which roughly coincides with the break in the radio luminosity function of star-forming galaxies (cfr Magliocchetti et al. 2002; Mauch \& Sadler 2007). Beyond this value,  their  luminosity function steeply declines, and the contribution of star-forming galaxies to the total radio population is drastically reduced to a negligible percentage. The same trend is true at higher redshifts, and since the radio luminosity function of star-forming galaxies drops off in a much steeper way than that of AGN at all $z$, we expect the chances of contamination in the two samples to be quite low.

Radio powers for our sample of 1537 radio-selected sources endowed with a redshift estimate have been calculated according to the relation:
\begin{eqnarray}
\rm P_{1.4 \rm GHz}=\rm F_{1.4 \rm GHz} D^2 (1+z)^{3+\alpha},
\end{eqnarray}
where the result is in [W Hz sr$^{-1}$] units, D is the angular diameter distance and $\alpha$ is the spectral index of the radio emission ($\rm F(\nu)\propto \nu^{-\alpha}$). 
Since we do not have estimates for this latter quantity, we adopted the average value $\alpha=0.7$ found for similar surveys (e.g. Randall et al. 2012 and references therein) both for star-forming galaxies and for AGN emission. Such an assumption should not be too far from the truth as 1) radio sources in the COSMOS-VLA sample are faint, therefore the chances of finding a large number of bright, flat spectrum AGN are  rather thin and 2) recent results report values  $\alpha\simeq 0.7$ also for star-forming galaxies at $z\simeq 2$ (Ibar et al. 2010), similar to what found locally for the same population (Condon 1992).

We then distinguished between AGN-powered galaxies and star-forming galaxies by means of equation (\ref{eq:P}) for $z\le 1.8$ and by fixing $\rm Log_{10}P_{\rm cross}(z)=23.5$ [W Hz sr$^{-1}$ ] at higher redshifts (cfr McAlpine, Jarvis \& Bonfield 2013). 
This procedure identifies 1026 sources (corresponding to $\sim 67$\% of the parent sample  endowed with a redshift determination) as star-forming galaxies and 482 as AGN. Note that, due to the adopted selection criteria and thanks to the great depth of the COSMOS-VLA survey, the AGN sample is {\it complete} with respect to radio selection at all redshifts $z\simlt 3.5$ ($F_{1.4 \rm GHz}\ge 0.06$ mJy corresponds to $\rm Log_{10}P \ge 23.5$ [W Hz sr$^{-1}$] at all $z\le 3.5$), i.e. the considered sample of COSMOS-VLA sources endowed with a redshift determination includes  {\it all} radio-emitting AGN, except for a small handful of sources (less than 10, cfr Figure 5) at the highest, $z>3.5$,  redshifts. 

Their distribution as a function of 1.4 GHz flux is presented in the top panel of Figure \ref{fig:idsvsstype} 
by the solid line, while the dashed line represents star-forming galaxies as singled out with our method. As expected, star-forming galaxies make up for the large majority of the radio sample at fluxes $\rm F_{1.4 GHz}\simlt 0.2$ mJy, while AGN dominate for  $\rm F_{1.4 GHz}\simgt 0.4$ mJy. 
The bottom panel of Figure \ref{fig:idsvsstype} instead represents the fraction of COSMOS-VLA radio galaxies with measured redshifts which also have a counterpart either at 100$\mu$m or at 160$\mu$m from the PEP survey (175 AGN, corresponding to $\sim 36$\% of the parent FIR-matched sample, and 657 star-forming galaxies). 
Again as expected, almost all star-forming galaxies (indicated by the dashed line) are associated to a FIR counterpart, the percentage decrement towards low radio fluxes only being due to the fact that the radio survey probes deeper fluxes than
the FIR sampling. On the other hand, the percentage of AGN which are also FIR emitters is about $40$\%, roughly constant with radio flux, at least up to $\rm F_{1.4 GHz}\simeq 3$ mJy

Figure \ref{fig:idsvsztype} illustrates the redshift distributions of the two populations of AGN (left-hand panel) and star-forming galaxies (right-hand panel). The solid histograms are for all sources, irrespective of their FIR emission, while the dashed lines indicate those objects which are also detected in the PEP survey. The two redshift distributions are noticeably different: in fact, while that referring to star-forming galaxies monotonically decreases from the lowest to the highest redshifts, the AGN distribution presents two distinct peaks, a first one at around $z\sim 0.9$, and a second one between $z\sim 2$ and $z\sim 3$. We stress that since the considered AGN sample is complete with respect to radio selection at all $z\simlt 3.5$, the solid histogram represents the redshift distribution of the {\it whole} population of radio-emitting AGN, except at the highest redshifts probed by our analysis.
 
Although rescaled, the distributions of the sub-populations of FIR emitters also follow a similar trend, both for star-forming galaxies and for AGN. The two lower panels  represent the ratio between the two radio-selected populations and the sub-sets of FIR emitters. These ratios are approximately constant: the percentage of FIR emitters amongst radio-selected AGN is about 30-40 \% at all redshifts, while that for radio-selected star-forming galaxies is constant and approximately equal to about 50\% in the redshift range  
$0.5 \simlt z\simlt 2.6$. Beyond this value, the limited number of sources does not allow to draw any conclusion on their behaviour.

\section{The Infrared properties of radio-selected AGN}

\begin{figure}
\includegraphics[scale=0.4]{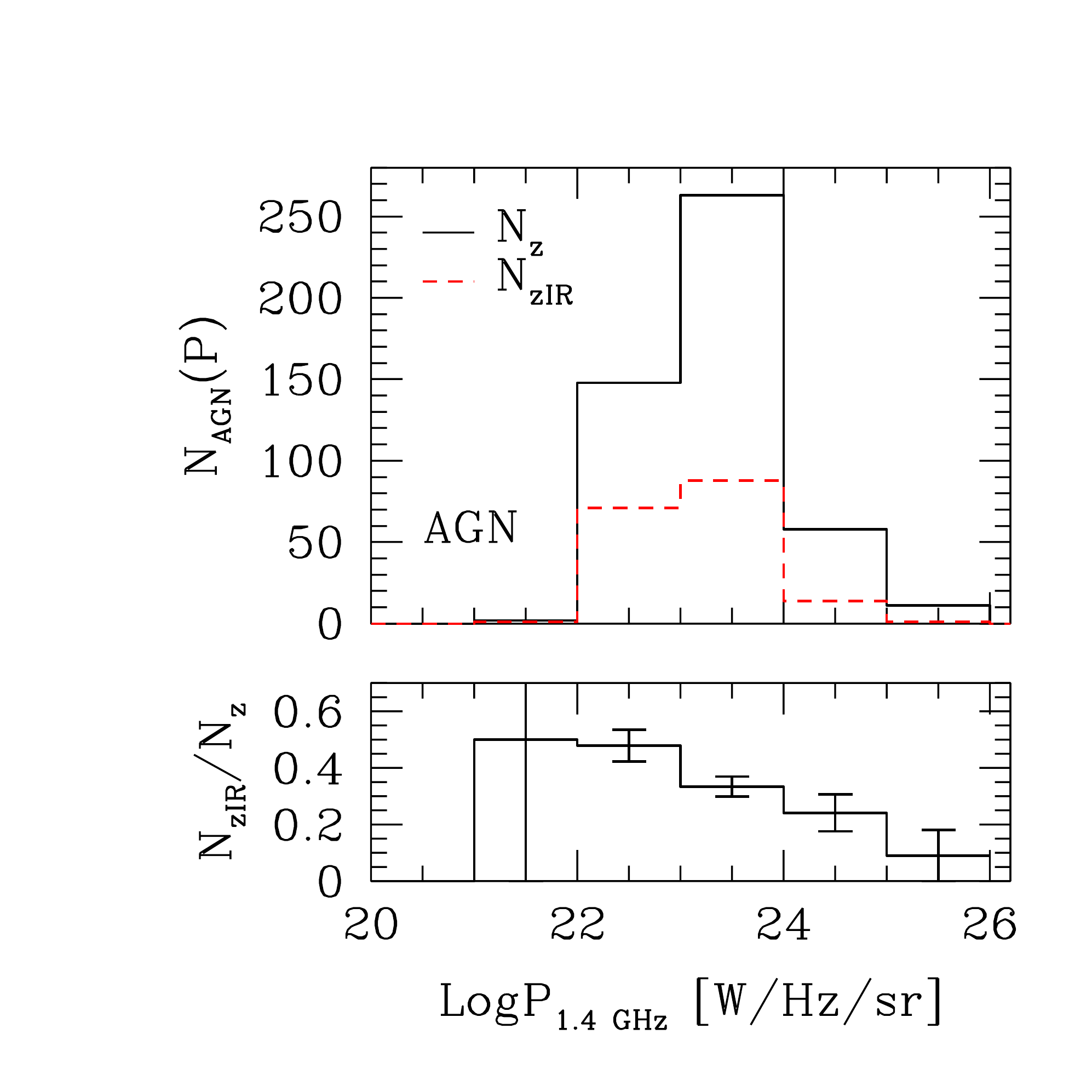}
\caption{Distribution of radio powers for F$_{\rm 1.4 GHz}\ge 0.06$ mJy COSMOS-VLA AGN. The solid line represents all sources, irrespective of their FIR emission, while the dashed line indicates those objects which are also FIR emitters. The bottom panel highlights the ratio between these two latter quantities. Errorbars correspond to Poissonian estimates.
\label{fig:idsvsPtype}}
\end{figure}

\begin{figure}
\begin{center}
%\vspace{4cm}  % amount of vertical space needed
\includegraphics[scale=0.4]{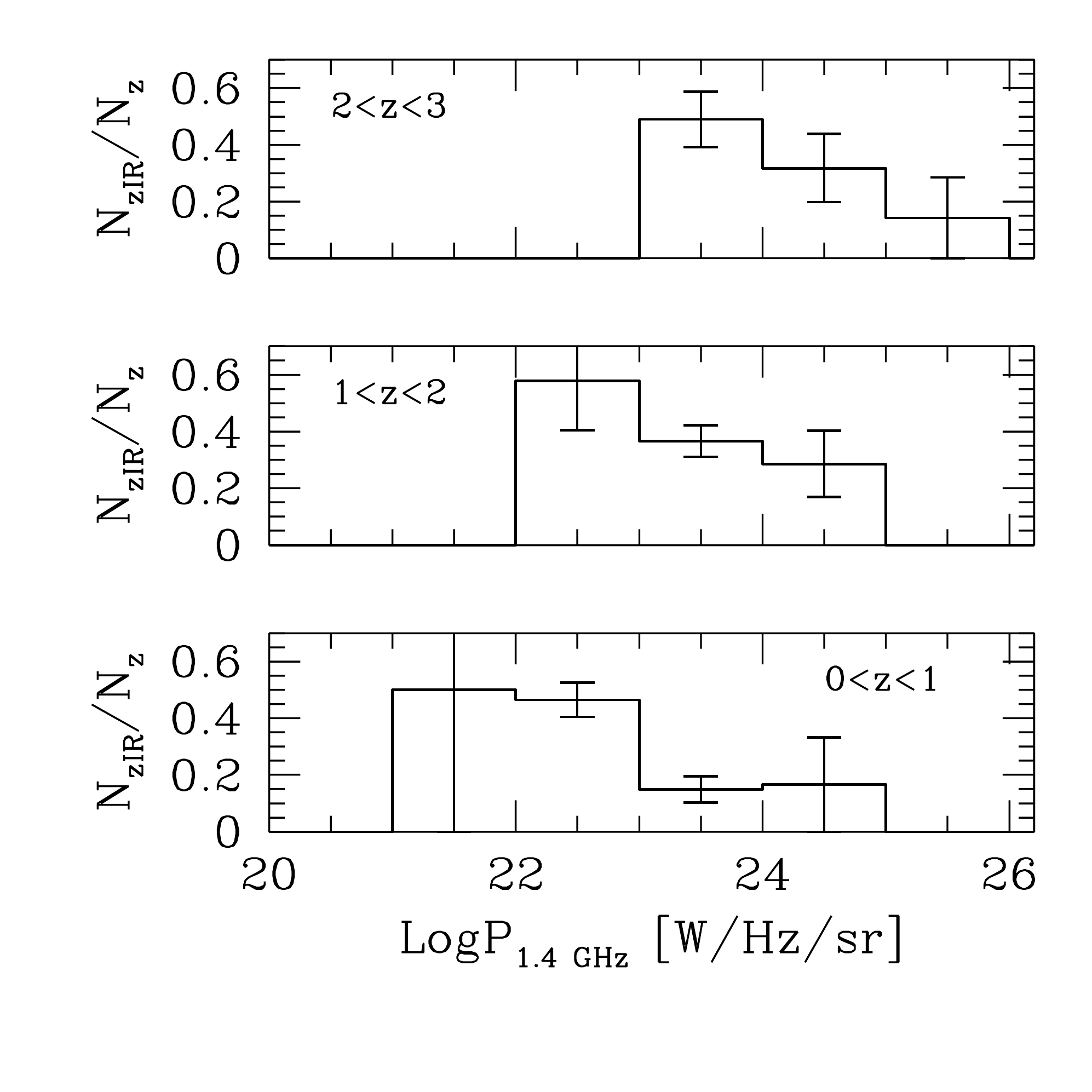}
\caption{Fraction of radio-selected AGN with FIR emission as a function of radio power in three different redshift bins. Errorbars correspond to Poissonian estimates.
\label{fig:idsvsPztype}}
\end{center}
\end{figure}

The distribution of radio powers for COSMOS-VLA AGN is presented in Figure \ref{fig:idsvsPtype}. The solid line represents all AGN, while the dashed line is for those which are also detected in the PEP survey. The fraction of FIR emitters (illustrated at the bottom of Figure \ref{fig:idsvsPtype}) is found to monotonically decrease with radio power, from $\sim 50$\% at the lowest luminosities down to $\sim 10$\% or less for P$_{1.4 \rm GHz}\simgt 10^{25}$ [W Hz sr$^{-1}$]. This result implies that the probability for a radio-selected AGN to emit in the FIR is  strongly dependent on its radio luminosity, the more luminous the source the slimmer the chances to have a detection at FIR wavelengths.

\begin{figure}
\begin{center}
%\vspace{4cm}  % amount of vertical space needed
\includegraphics[scale=0.4]{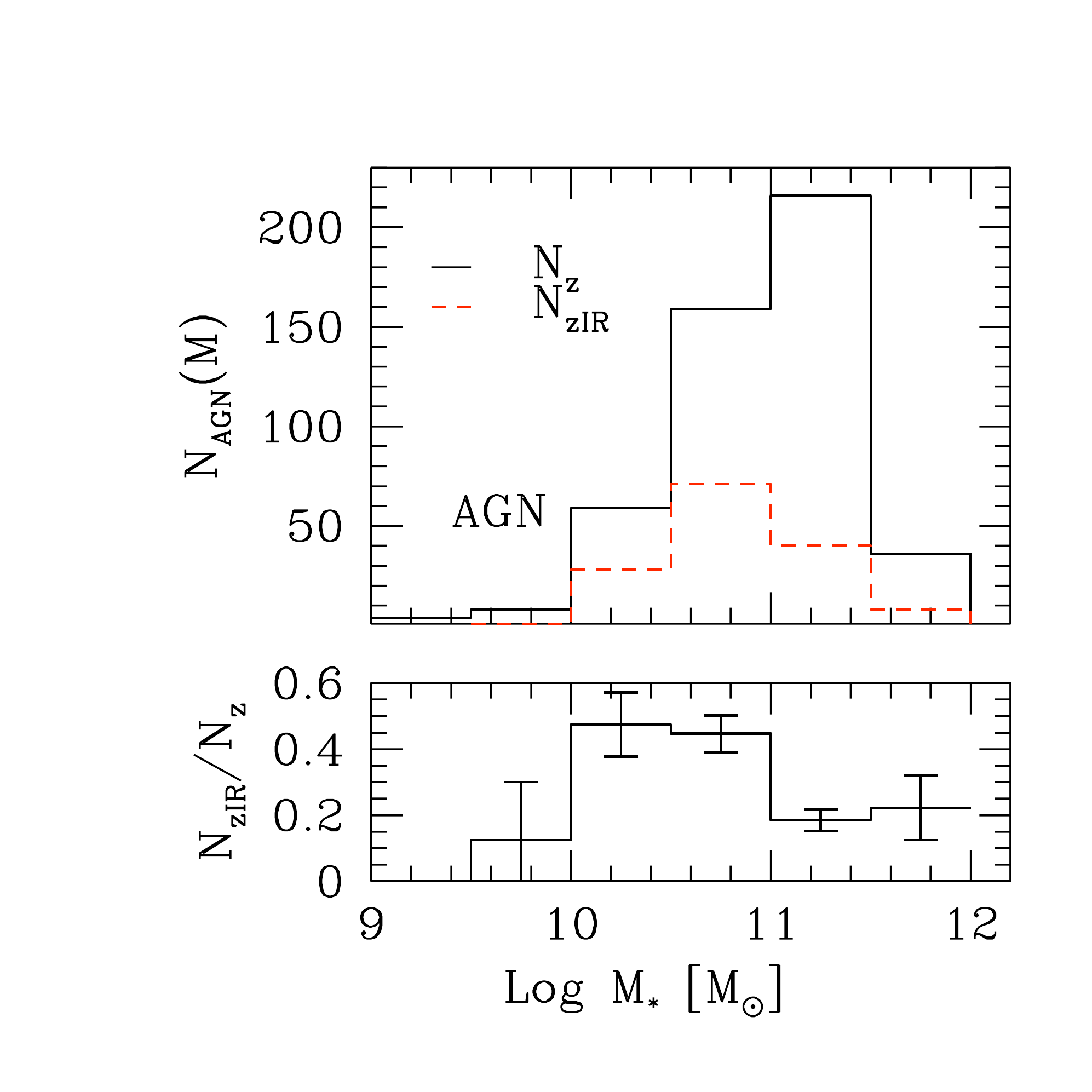}
\caption{Stellar mass distributions for F$_{\rm 1.4 GHz}\ge 0.06$ mJy  COSMOS-VLA AGN. The solid line represents all sources, irrespective of their FIR emission, while the dashed line indicates those objects which are also FIR emitters. The bottom panel highlights the ratio between these two latter quantities. Errorbars correspond to Poissonian estimates.
\label{fig:idsvsmasstype}}
\end{center}
\end{figure}

\begin{figure}
\begin{center}
%\vspace{4cm}  % amount of vertical space needed
\includegraphics[scale=0.4]{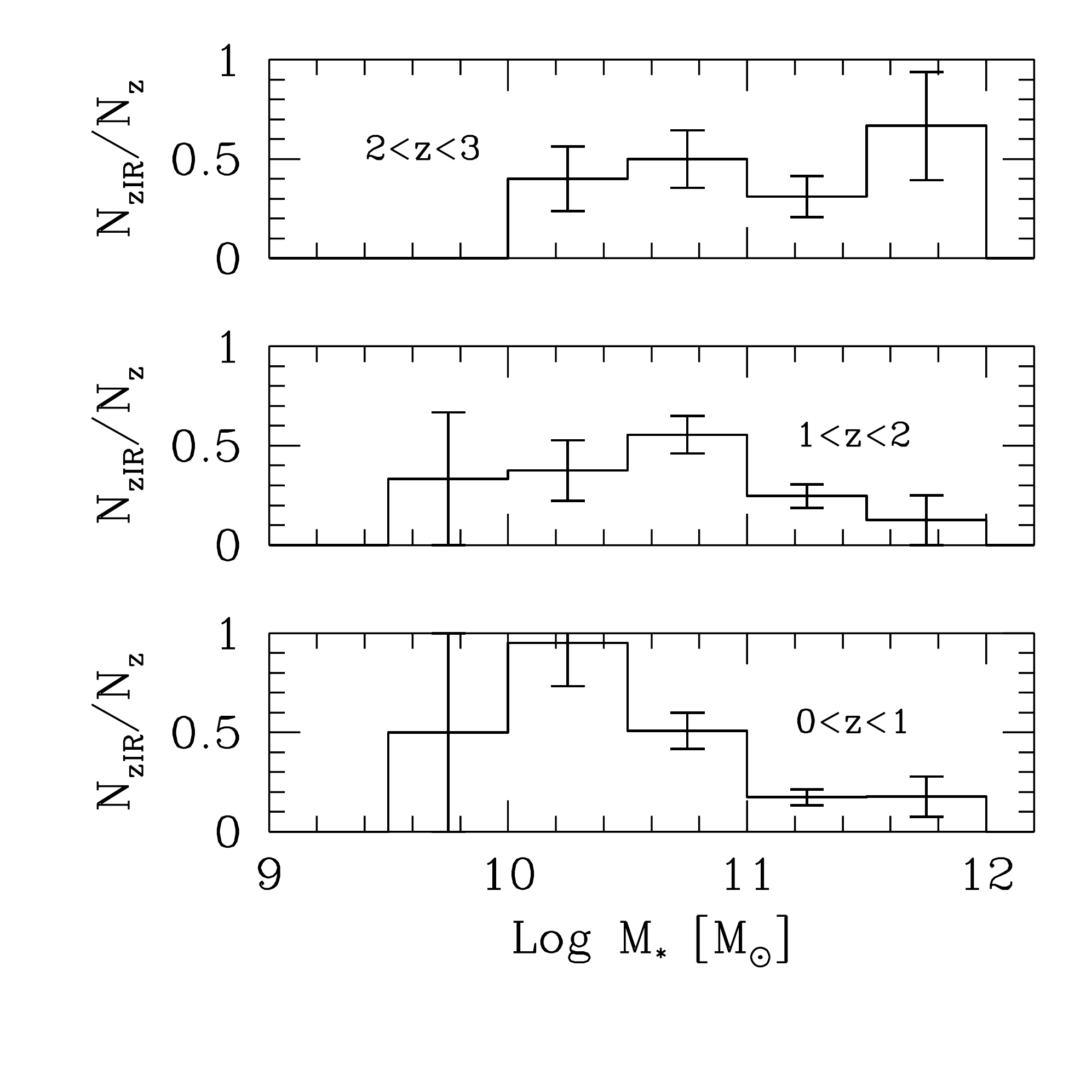}
\caption{Fraction of radio-selected AGN with FIR emission as a function of stellar mass in three different redshift bins. Errorbars correspond to Poissonian estimates.
\label{fig:idsvsmassztype}}
\end{center}
\end{figure}

However, this is not the whole story. A better insight on the above result can be gained by splitting the AGN sample into different redshift bins. This is presented in Figure \ref{fig:idsvsPztype} which clearly shows that, while at all $z$ it is still true that powerful radio AGN are less likely to be FIR emitters, this however happens at different radio luminosities in the different redshift ranges. 
%In more detail, while for $z< 1$the fraction of FIR emitters drops  to an almost negligible percentage above P$_{1.4 \rm GHz}\simgt 10^{23}$ [W Hz sr$^{-1}$], the same is not found at higher redshifts. 
In fact, we find that the chances for an AGN of fixed radio luminosity to be a FIR emitter sensibly increase when moving from the lowest to the highest redshifts probed by our analysis. So, for instance, a source with P$_{1.4 \rm GHz}\simeq 10^{23}$ [W Hz sr$^{-1}$] will only have 
a $\sim 20$\% probability of being a FIR emitter at $z\le1$, while this percentage rises to $\sim 40$\% in the redshift range $z=[1-2]$, up to $\sim 50-60$\% for $z=[2-3]$. \\
From the above analysis, we can then conclude that  {\it the probability for a radio-selected AGN to be active in the FIR is both a function of radio power and redshift, whereby powerful sources are more likely to emit at FIR wavelengths at higher redshifts}.

The COSMOS photometric catalogue also provides stellar mass estimates for the majority of the sources endowed with a redshift determination. We used this information to compute the mass distribution of AGN drawn from the $\rm F_{1.4 GHz}\ge 0.06$ mJy COSMOS-VLA sample as previously explained. The results are plotted in  Figure \ref{fig:idsvsmasstype} where, once again, the solid histogram refers to all radio-selected AGN, irrespective of their FIR emission. 

The first interesting result we find is that radio-selected AGN are hosted by very massive galaxies. The low-mass, $\rm M_*\simlt 10^{10} M_\odot$, tail of the distribution is practically absent and the distribution itself is quite narrowly centered around $\rm M_*\sim 10^{11.2} M_\odot$. 
This is in agreement with results from the clustering of radio-active AGN which show them to be hosted by very massive, $\rm M_h\simgt 10^{13.5} M_\odot$ dark matter halos (e.g. Magliocchetti et al. 2004: Fine et al. 2011).

The distribution of masses of COSMOS-VLA AGN with a PEP counterpart (presented in Figure  \ref{fig:idsvsmasstype} by the dashed histogram) is also quite interesting. 
First of all, the sub-population of FIR emitters tends to be hosted by smaller galaxies. Their distribution in fact peaks at $\rm M_*\sim 10^{10.7} M_\odot$, vs the $\rm M_*\sim 10^{11.2} M_\odot$ value found for the whole population of radio-selected AGN. 
Furthermore, we find that about 50\% of AGN has a FIR counterpart only in the very narrow $[10^{10}-10^{11}] \rm M_\odot$ range, while this percentage is drastically reduced to values below $\sim 20$\% both at  lower and higher masses. In other words, it seems that {\it there is a preferential mass scale for which radio AGN can also be active at FIR wavelengths}.

As in the previous case, more information on the above finding can be gathered by splitting the AGN sample into redshift bins and re-estimating the mass distribution as a function of $z$. The results are presented in Figure 
\ref{fig:idsvsmassztype}, which clearly shows that FIR emitters are preferentially found in smaller, $\rm M_*\simlt 10^{11} M_\odot$, galaxies only for $z\le 2$. There appears not to be any specific trend at higher redshifts, whereby the chances for a radio-selected AGN to also be a FIR emitter are independent of the mass of its host galaxy. 

\section{The origin of FIR emission in radio-selected AGN}

\begin{figure*}
\includegraphics[scale=0.4]{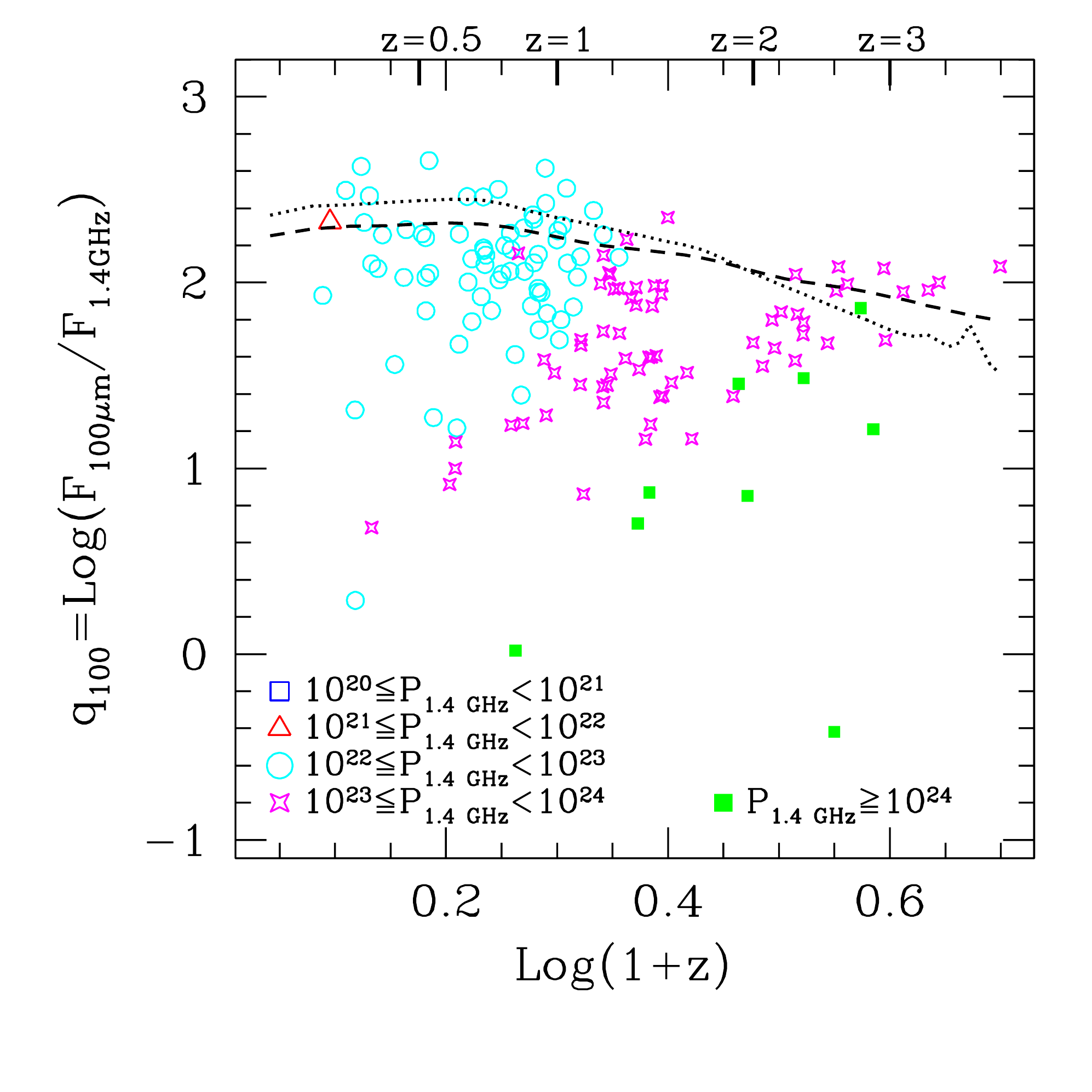}
\includegraphics[scale=0.4]{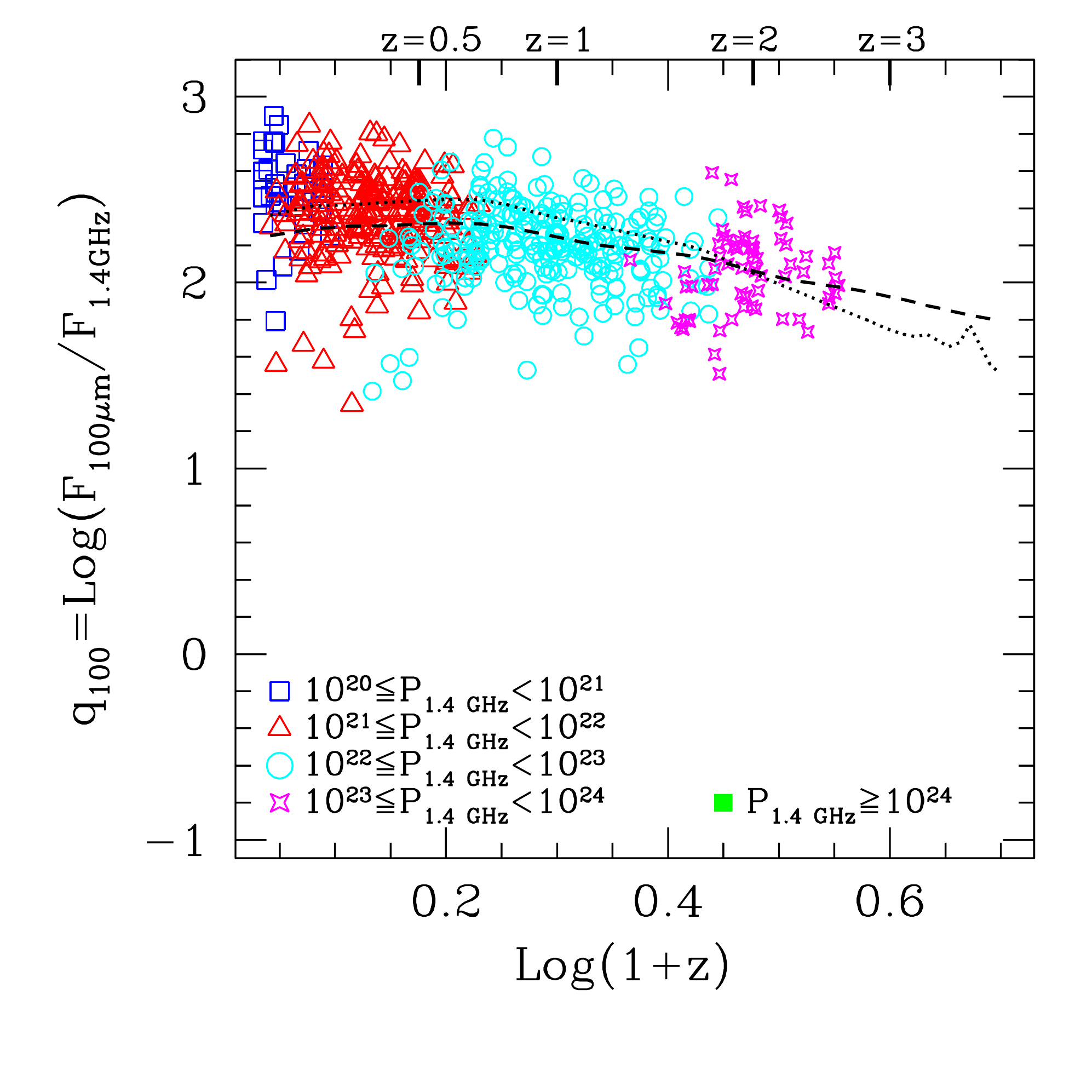}
\caption{Ratio between F$_{100 \rm \mu m}$ and F$_{1.4 \rm GHz}$ fluxes as a function of redshift for radio-selected sources with a FIR counterpart in the PEP survey classified as AGN (left-hand panel) or star-forming galaxies (right-hand panel). Different symbols correspond to different intervals for the radio power (measured in [W Hz sr$^{-1}$]).The dashed line represents the trend obtained for the SED of M82, while the dotted one corresponds to Arp220 (see text for details).
\label{fig:q100}}
\end{figure*}

\begin{figure*}
\includegraphics[scale=0.4]{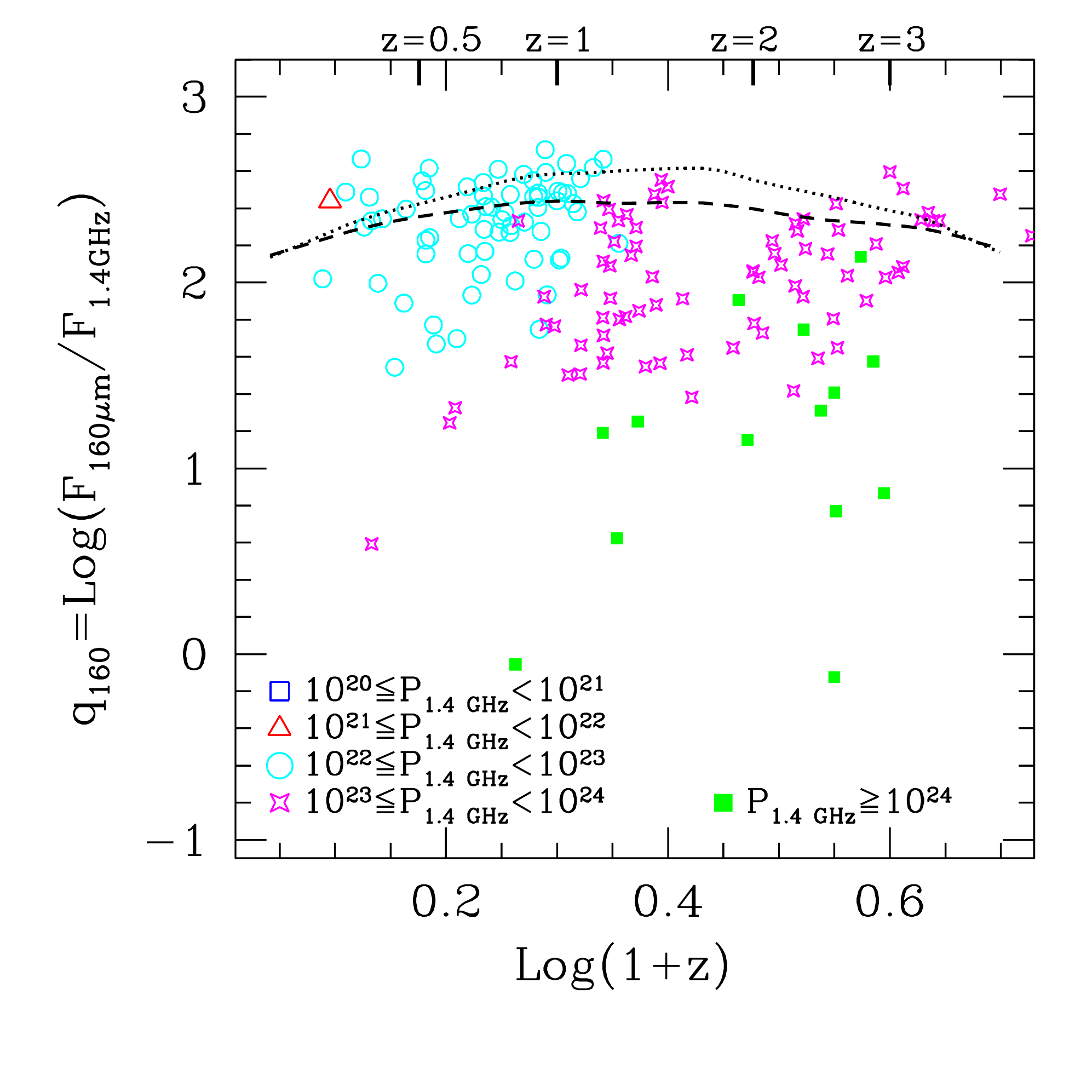}
\includegraphics[scale=0.4]{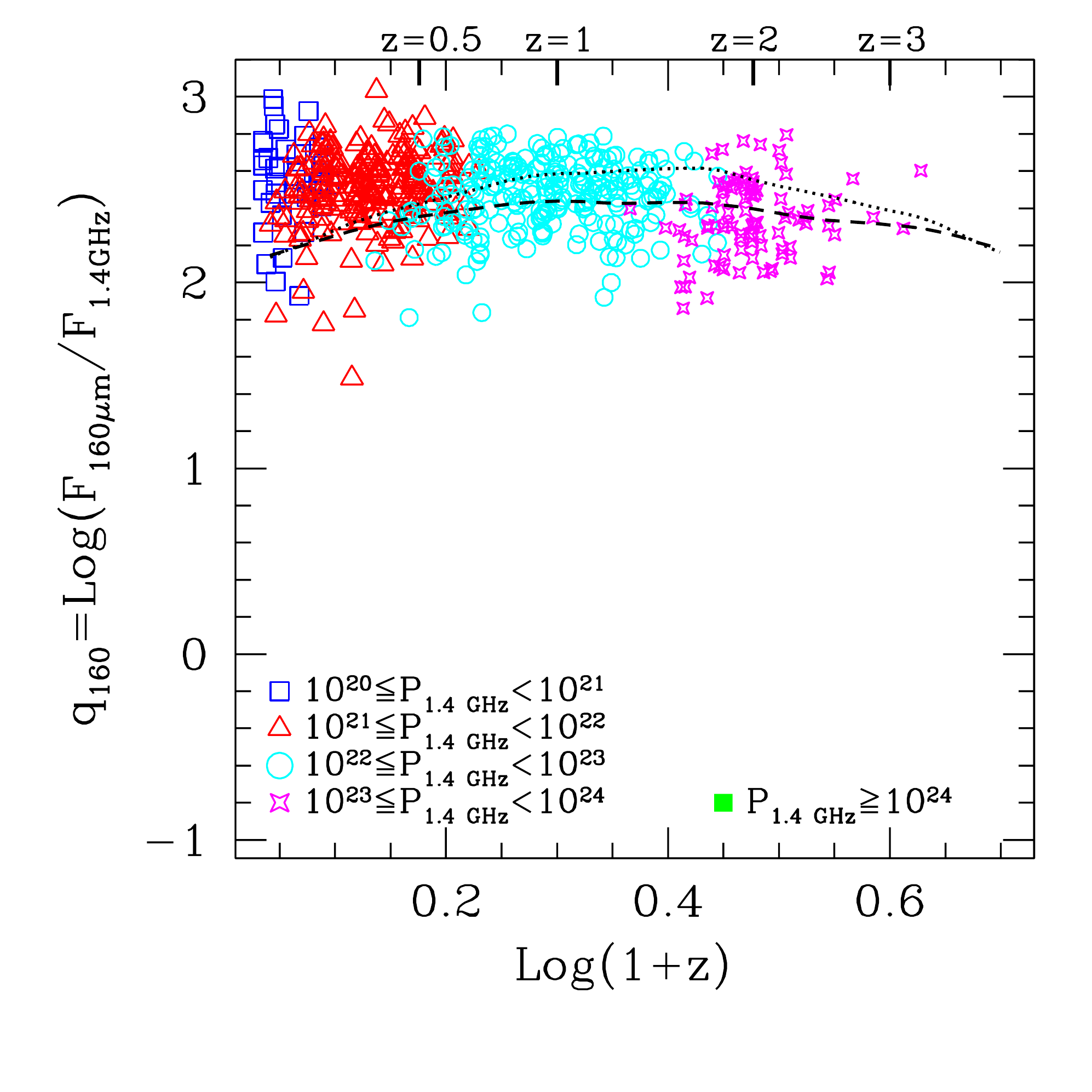}
\caption{Ratio between F$_{160 \rm \mu m}$ and F$_{1.4 \rm GHz}$ fluxes as a function of redshift for radio-selected sources with a FIR counterpart in the PEP survey classified as AGN (left-hand panel) or star-forming galaxies (right-hand panel). Different symbols correspond to different intervals for the radio power  (measured in [W Hz sr$^{-1}$]). The dashed line represents the trend obtained for the SED of M82, while the dotted one corresponds to Arp220 (see text for details).
\label{fig:q160}}
\end{figure*}

\begin{figure*}
\includegraphics[scale=0.4]{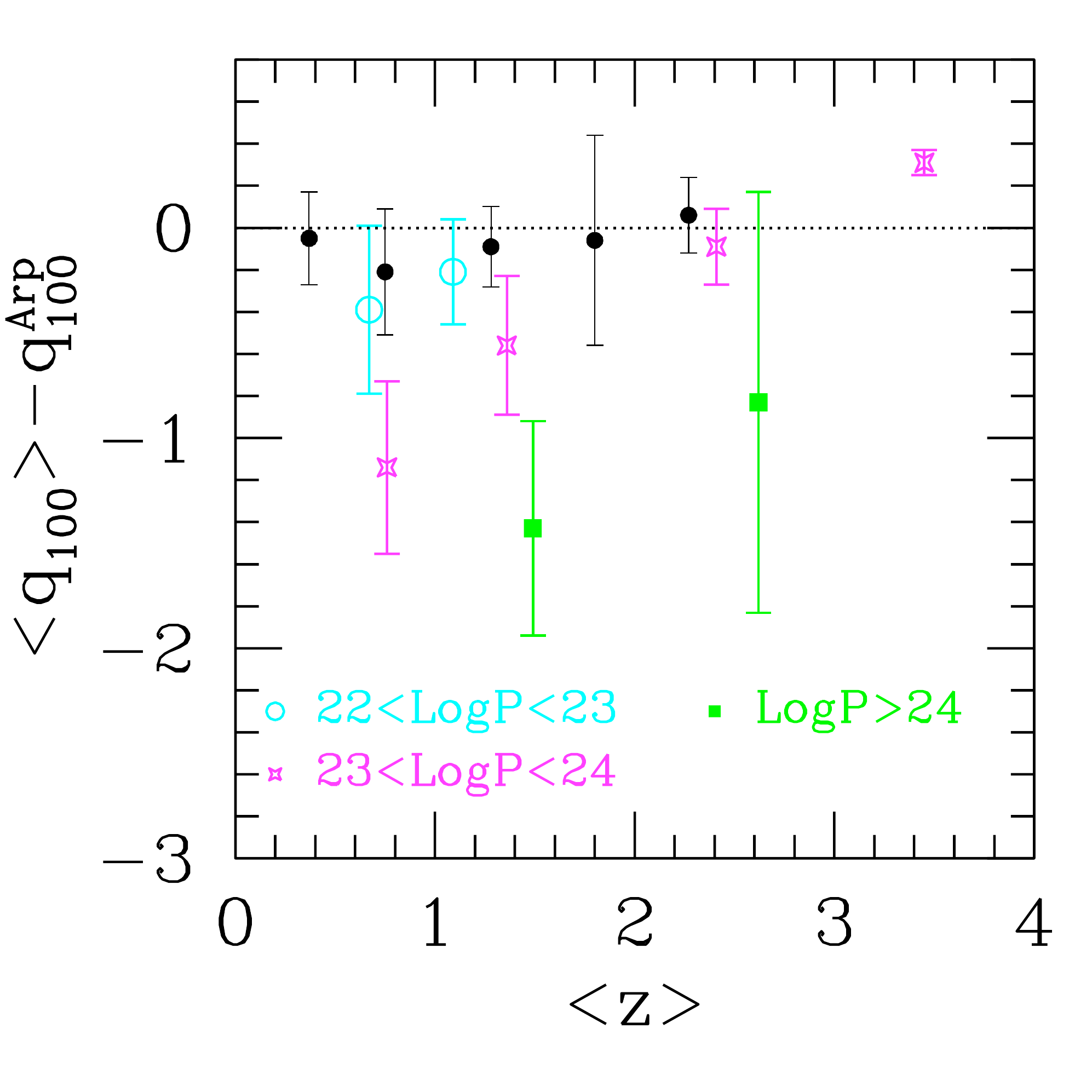}
\includegraphics[scale=0.4]{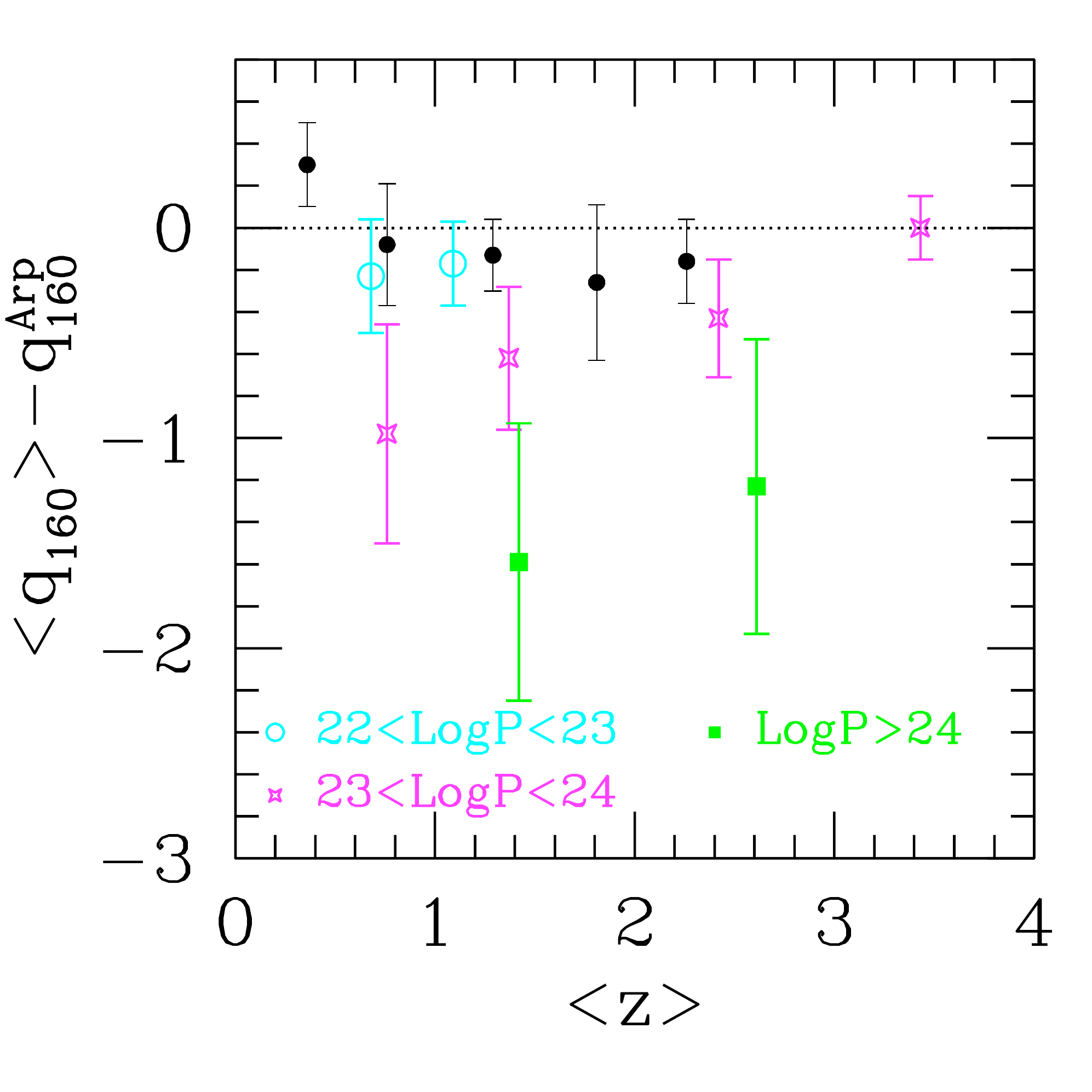}
\caption{Redshift dependence of the difference between the average values of q$_{100}$ (left-hand panel) and q$_{160}$ (right-hand panel) for  F$_{1.4 \rm GHz}\ge 0.06$ mJy AGN and the same quantities obtained for an Arp220-like SED. Different symbols correspond to different ranges in radio luminosity. Errorbars are derived from Poissonian estimates. The plotted redshifts correspond to the average values of the AGN population (respectively with a counterpart at 100$\mu$m and 160$\mu$m) in the redshift ranges $z=[0-1]$, $z=[1-2]$, $z=[2-3]$ and $z=[3-4]$. For a comparison, the filled dots represent  the distribution of residuals 
$<{\rm q}_{100}>-{\rm q}_{100}^{\rm Arp}$ (left-hand panel) and $<{\rm q}_{160}>-{\rm q}_{160}^{\rm Arp}$ (right-hand panel)
derived for the subset of star-forming galaxies.
\label{fig:qav}}
\end{figure*}

\begin{figure*}
\includegraphics[scale=0.4]{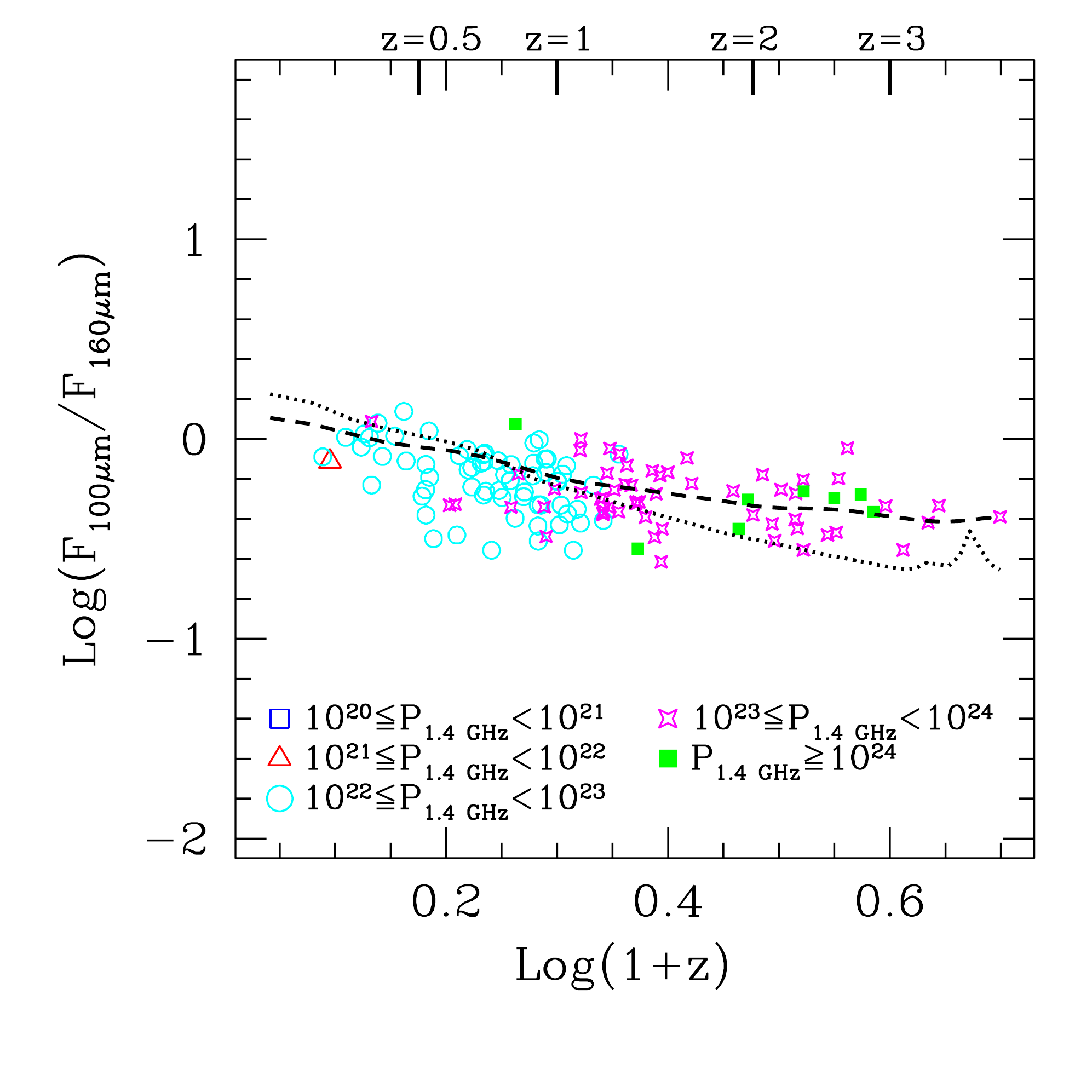}
\includegraphics[scale=0.4]{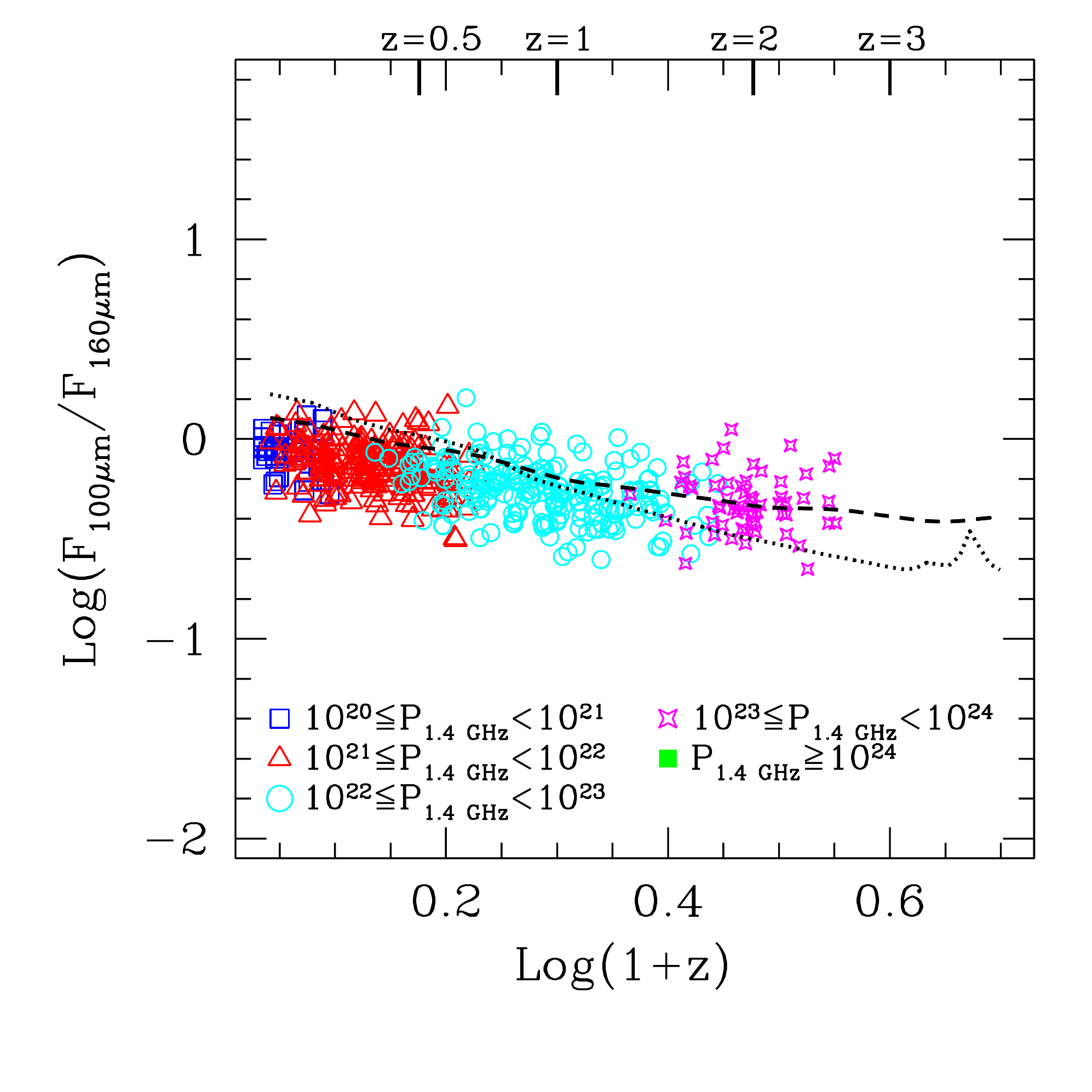}
\caption{Ratio between F$_{100 \rm \mu m}$ and F$_{160 \rm \mu m}$ fluxes as a function of redshift for radio-selected sources with a FIR counterpart in the PEP survey classified as AGN (left-hand panel) or star-forming galaxies (right-hand panel). Different symbols correspond to different intervals for the radio power  (measured in [W Hz sr$^{-1}$]). The dashed line represents the trend obtained for the SED of M82, while the dotted one corresponds to Arp220 (see text for details).
\label{fig:q100_160}}
\end{figure*}

In the previous Section we have obtained two important results. The first one is that the probability for a radio-selected AGN to also be a FIR emitter is a strong function of both radio luminosity and redshift. Such a probability in fact monotonically decreases for brighter sources, but the radio luminosity at which this effect starts appearing increases as we move to higher $z$'s. We have also found that, at all redshifts $\simlt 2$, there is a typical (narrow) galaxy mass range M$_*\simeq [10^{10}-10^{11}] \rm M_\odot$ which maximizes the chances for an AGN to also emit at FIR wavelengths. 
However, the above results do not shed any light on what is (are) the physical process(es) responsible for FIR emission in radio-active AGN. 

In order to answer this question, we have then considered the photometrical properties of such sources. More precisely, we have analysed the distribution of infrared and radio fluxes and compared them with those exhibited by the sub-population of star-forming galaxies, selected from the COSMOS-VLA sample by applying the prescription given in \S 3. 
Figures \ref{fig:q100} and \ref{fig:q160} present the distribution of flux ratios $\rm q_{100}=log_{10}F_{100 \mu m}/log_{10}F_{1.4 GHz}$ and $\rm q_{160}=log_{10}F_{160 \mu m}/log_{10}F_{1.4 GHz}$ plotted as a function of look-back time for radio-selected sources of different radio luminosities. The left-hand panels show the case for AGN, while the right-hand panels refer to star-forming galaxies (a more exhaustive work on the evolution of the FIR-to-radio relation based on PEP data is presented in Magnelli et al. in preparation). The dashed lines  reproduce the q$_{100}$ and q$_{160}$ ratios obtained from the spectral energy distribution (SED) of  M82, while the dotted ones are those expected for an Arp220-like galaxy. As it is clearly visible, the distributions obtained for the two different populations are remarkably different: star-forming galaxies smoothly sit on the template curves that are appropriate for this population while,  at all radio luminosities, the large majority of  AGNs  presents boosted radio activity which does not appear to be correlated with their FIR emission.

More information can be gathered from Figures \ref{fig:q100} and \ref{fig:q160}  which show some trend for
 the q$_{100}$ and q$_{160}$ values of AGN of fixed radio luminosity  to increase  and approach those indicated by the template star-forming SEDs as we move up in redshifts. 
This is better seen in Figure \ref{fig:qav} which represents the residuals $\Delta$q between the average values of q$_{100}$ (left-hand panel) and q$_{160}$ (right-hand panel) obtained for radio-selected AGN and the corresponding Arp220 SED as a function of redshift for different radio luminosities. At all radio luminosities and in both the q$_{100}$ and q$_{160}$ cases, the residuals decrease and tend to the zero value (highlighted by the horizontal dotted lines) as one moves back in time. The effect is more pronounced for $23\le\rm log_{10} P_{1.4 GHz}\le 24$ [W Hz sr$^{-1}$] sources, mainly because this is the only case which covers the whole redshift range probed by our analysis and shows the best statistics,  but it can still be appreciated, although to a lower significance level, at the other radio luminosities. 
Note that, since the above analysis has been performed at fixed radio luminosities, this result implies that {\it the observed trend is solely due to an enhancement of FIR activity}. \\ As a comparison, we have also plotted the $<{\rm q}_{100}>-{\rm q}_{100}^{\rm Arp}$ and $<{\rm q}_{160}>-{\rm q}_{160}^{\rm Arp}$ residuals obtained for the sub-class of star-forming galaxies (filled dots with associated errorbars) at all radio luminosities. As expected, in this case the data does not show any dependency on either radio power or redshift and the photometric properties of these sources are perfectly compatible with an Arp220 SED.

Another point from Figure \ref{fig:qav} which is important to stress is  that the dependence observed at low redshifts between the quantities $\Delta$q  and radio luminosity  (brighter luminosities correspond to larger absolute values of $\Delta$q)
tends to be reduced as one moves to earlier epochs.
 By complementing these findings with the results obtained in the previous Section, we can then conclude that 
{\it the fractional increment of FIR emitters amongst radio-selected AGN with look-back time is due to enhanced FIR activity of such sources at earlier epochs.} 
Furthermore,  the gradual reduction with $z$ of the $\Delta$q-to-radio luminosity correlation observed in the local universe  hints to a cosmological evolution, with AGN activity increasingly less effective at inhibiting FIR emission at the earlier epochs. 

The origin of FIR emission in radio-selected AGN can then be better understood by examining the distribution of their 
F$_{100 \rm \mu m}/\rm F_{160 \rm \mu m}$  colours and comparing them with those exhibited by the population of star-forming galaxies. This is done in Figure \ref{fig:q100_160}, which presents such colours as a function of redshift, for different intervals of radio luminosity. Once again, the dotted and dashed lines represent the FIR colours of an M82-like galaxy and of an Arp220-like galaxy. Radio-selected AGN present a distribution which is indistinguishable from that of star-forming galaxies. Both populations  tightly follow the trend indicated by the two template (star-forming) SEDs. Even more interestingly, this happens at all radio luminosities, even for the most powerful, $\rm log_{10} P_{1.4 GHz}\ge 24$ [W Hz sr$^{-1}$] sources, which do not show any deviation from the general trend. It is then straightforward to conclude that {\it FIR emission in radio-selected AGN is due to processes connected with ongoing star-formation within the host galaxy. At FIR levels, such processes are indistinguishable from those powering 'pure' star-forming galaxies.} 
And since star-forming processes also generate radio emission, another important conclusion of our analysis is that {\it the observed radio signal produced by radio-selected AGN which are also active at FIR wavelengths is the superposition of two contributions, one due to AGN accretion and one caused by stellar formation within the host galaxy. }

\begin{figure*}
\includegraphics[scale=0.4]{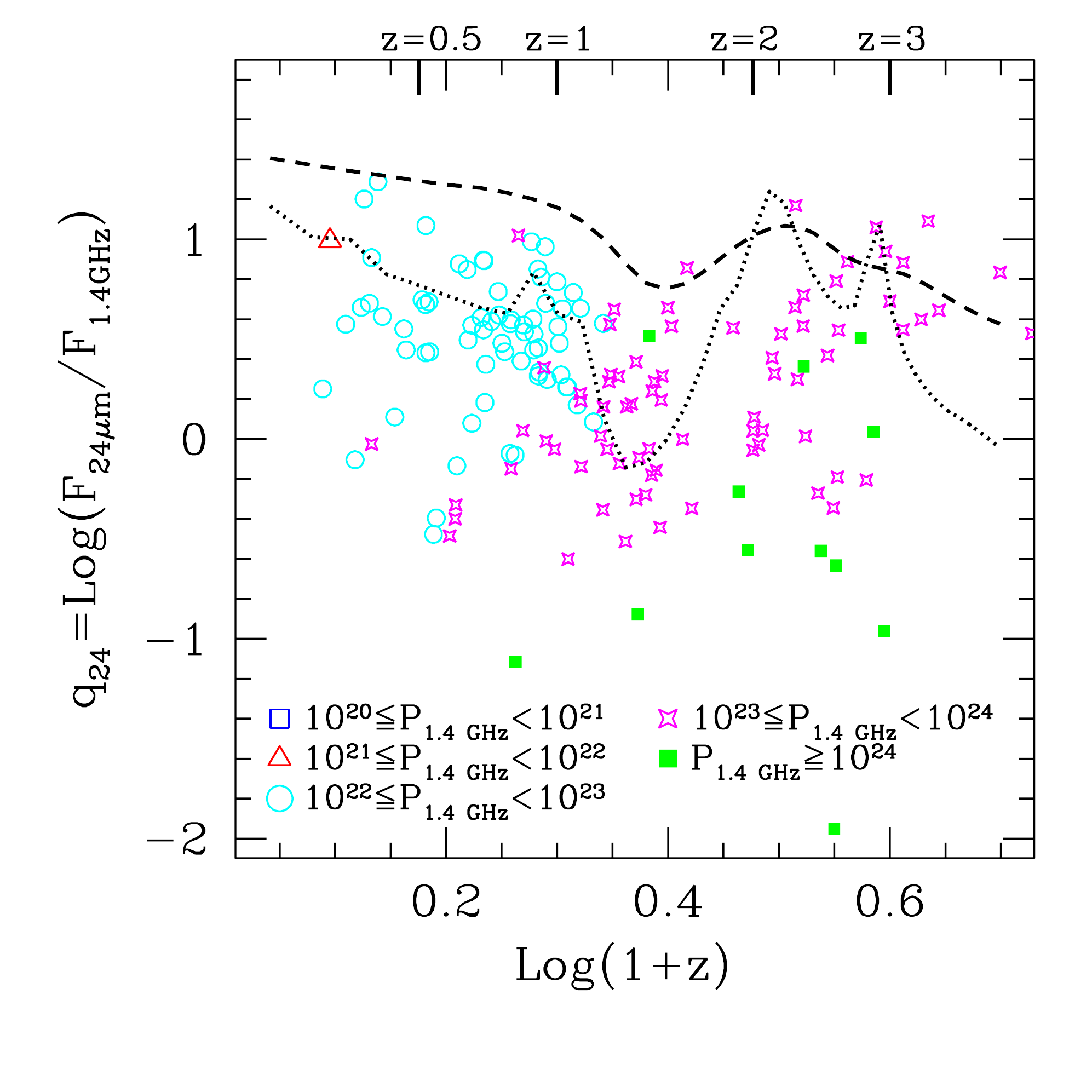}
\includegraphics[scale=0.4]{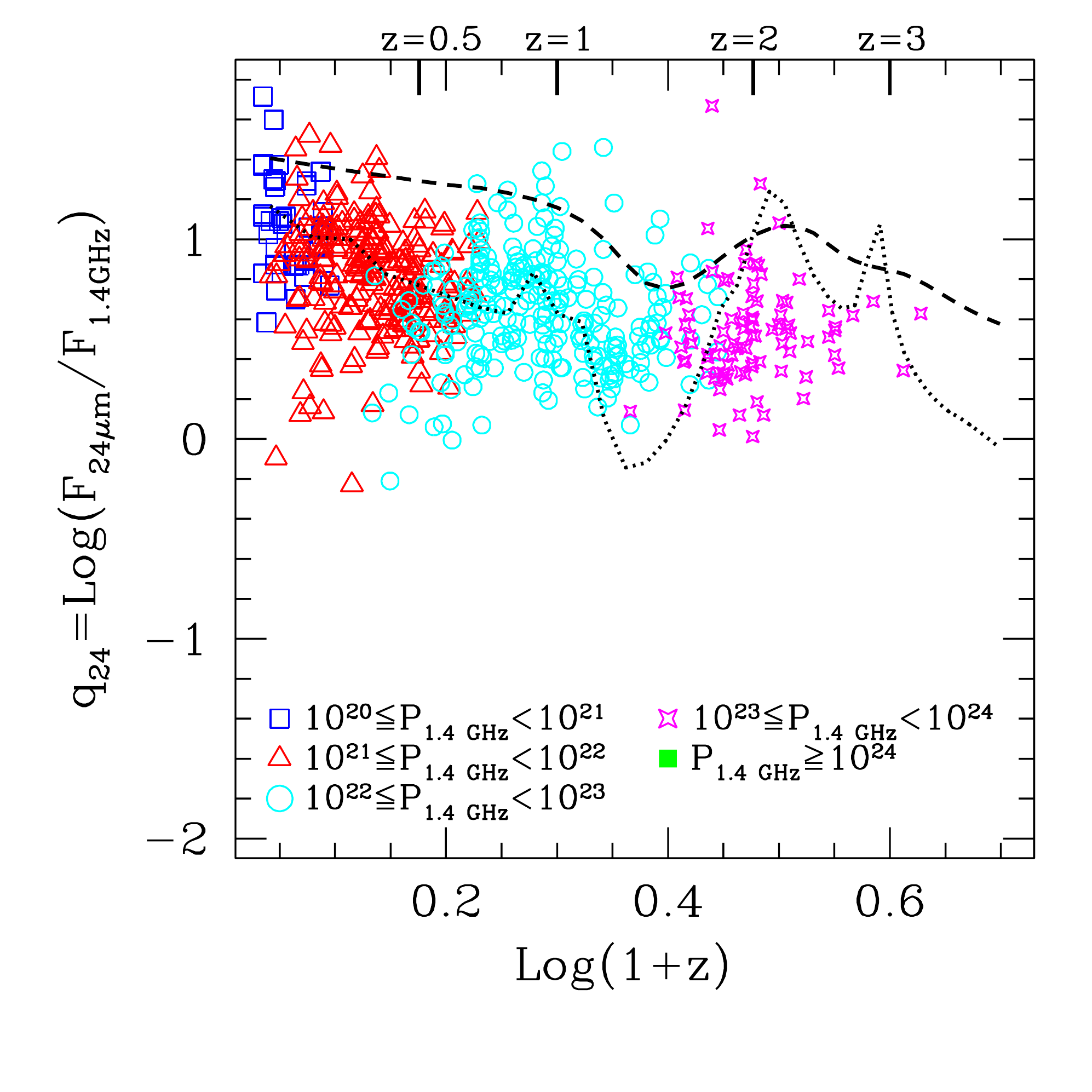}
\includegraphics[scale=0.4]{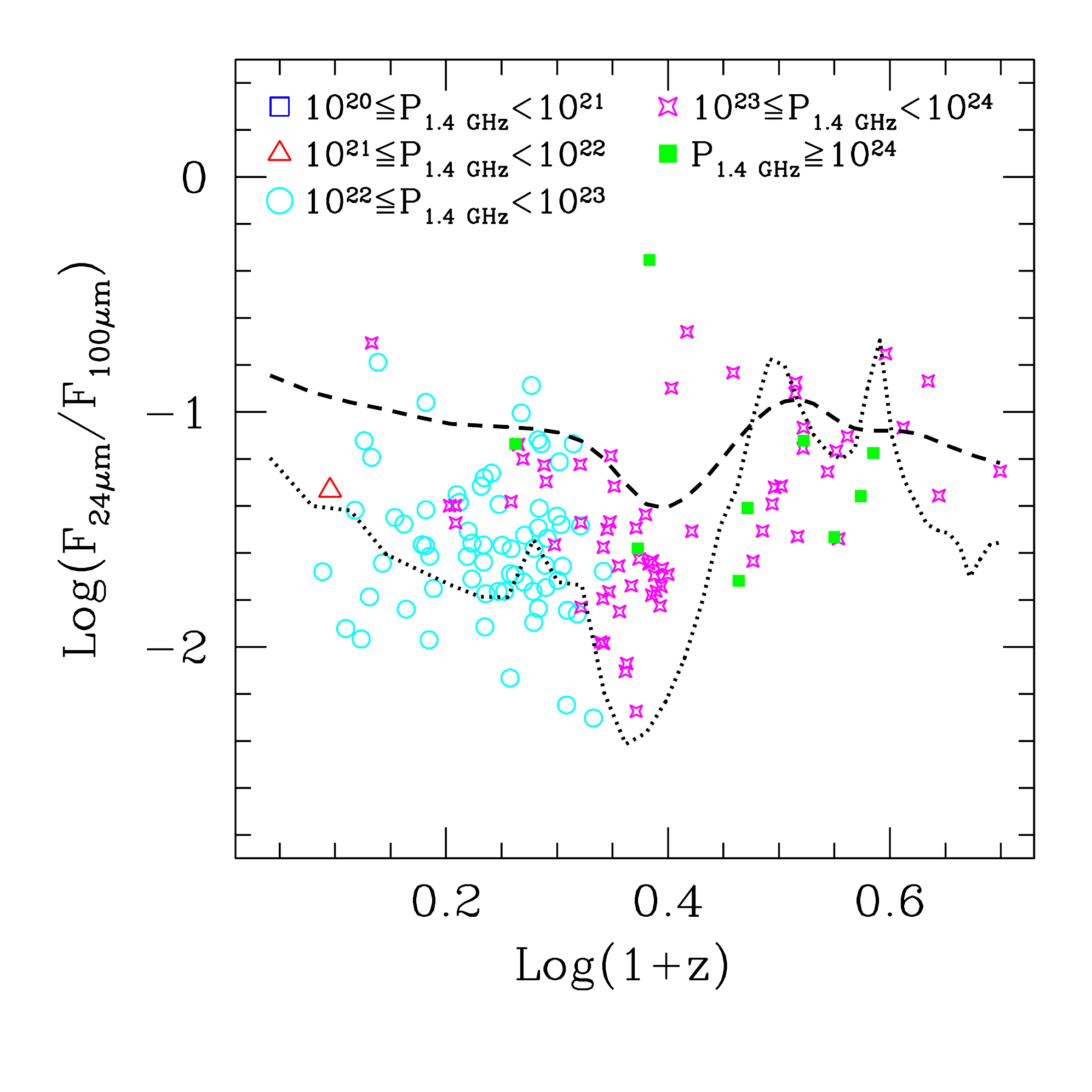}
\includegraphics[scale=0.4]{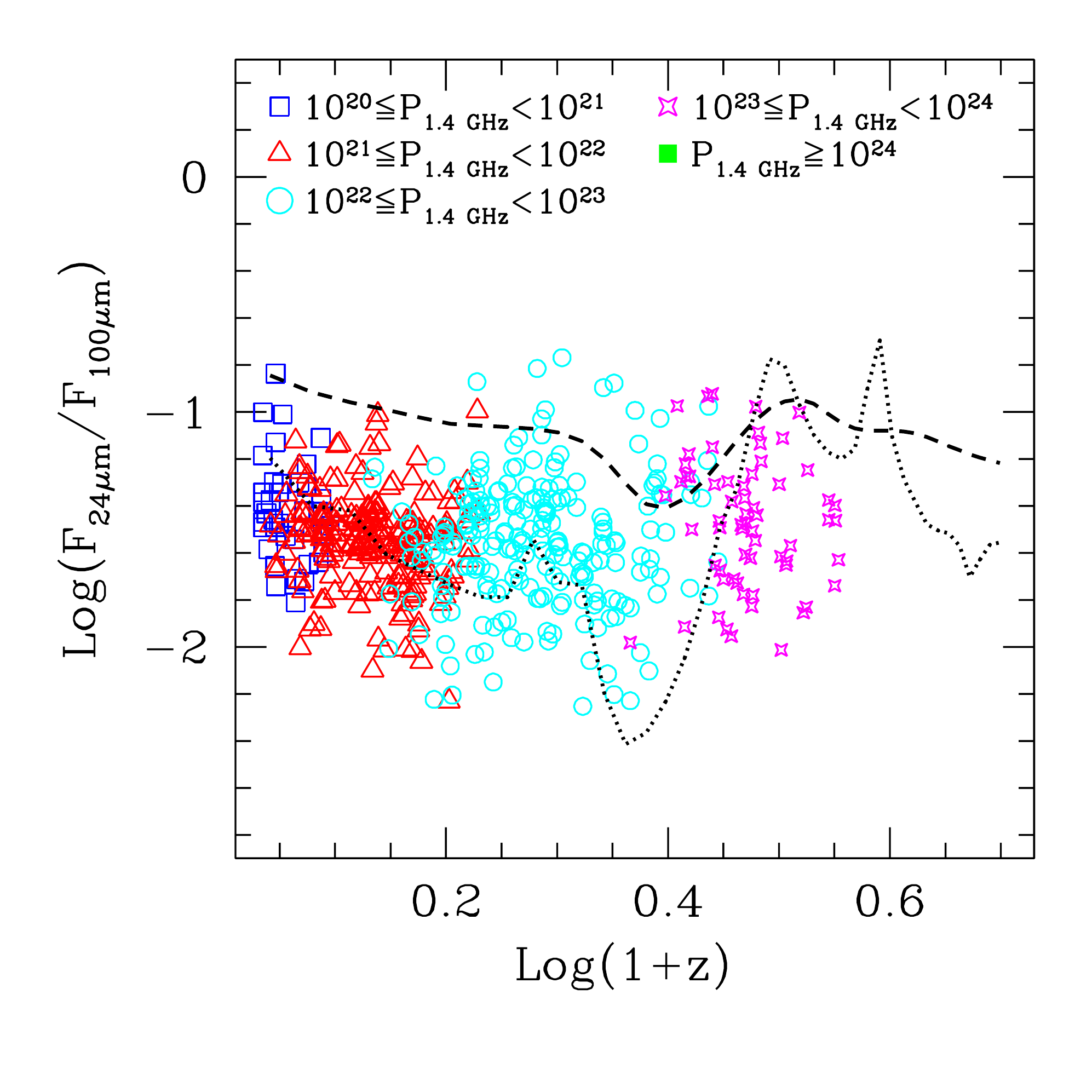}
\caption{F$_{24 \rm \mu m}$/F$_{1.4 \rm GHz}$ (top panels) and F$_{24 \rm \mu m}$/F$_{100\mu\rm m}$ colours (bottoms panels) as a function of redshift for radio-selected sources with a FIR counterpart in the PEP survey classified as AGN (left-hand panels) or star-forming galaxies (right-hand panels). Different symbols correspond to different intervals for the radio power  (measured in [W Hz sr$^{-1}$]).
The dashed lines represent the trend obtained for the SED of M82, while the dotted ones correspond to Arp220 (see text for details).
\label{fig:q24}}
\end{figure*}

The above results surely hold for the $\sim$35\% of the COSMOS-VLA sample of radio-selected AGN which also have a FIR counterpart from PEP. What about the rest of the AGN population? Are our conclusions still valid or do selection processes play an important biasing role?  In order to answer this question, first of all we recall that, for radio fluxes F$_{1.4\rm GHz}\ge 0.06$ mJy, the AGN selection we performed based on equation (\ref{eq:P}) ensures the sample is basically complete in radio selection at all available redshifts.
 And since the fraction of radio-selected AGN which have a FIR counterpart is roughly independent of both redshift and radio flux (cfr Figures 4 and 5), we expect that probing FIR fluxes that are fainter than those observed by the PEP survey would merely have the effect of increasing the fraction of FIR-detected radio AGN, while leaving all the conclusions of this paper unchanged. \\
 One might also wonder whether the findings of this work could be affected by contamination of the AGN sample with star-forming galaxies. 
To answer this concern, we first notice that, since the sample was selected only on the basis of radio emission and since the radio luminosity function of star-forming galaxies is much steeper than that of AGN at all z, chances of contamination based on selection for radio luminosities above a chosen threshold are very limited. 
It is true that, since the two populations have been originally divided by Mc Alpine et al. (2012) on the basis of their optical and near infrared SEDs, there might be cases in which an AGN appearing in the optical/NIR bands might not be active at radio wavelengths, so that the radio signal observed for that object has to be attributed to star-forming processes in the host galaxy rather than to accretion onto a black hole. These sources would result to follow the various q$_{\rm FIR}$ behaviours which are appropriate for star-forming galaxies, and in general show relatively low radio powers (crf \S 3). However we do not expect many of these cases. And, most importantly, none of the conclusions of this work would change if we remove these sources, since the findings we derive for our AGN sample are based on smooth trends throughout  the whole range of radio powers.

Lastly, we also considered 24$\mu$m fluxes and the distribution of MIR-to-radio, F$_{24 \rm \mu m}$/F$_{1.4 \rm GHz}$,  and MIR-to-FIR, F$_{24 \rm \mu m}$/F$_{100 \rm \mu m}$, colours as a function of redshift. 
This is done in Figure \ref{fig:q24}. As it was for the distribution of FIR colours, we find that there is not much of a difference in the behavior of AGN and star-forming galaxies.
The AGN distribution is more spread out and these objects still show some enhancement in their radio activity with respect to star-forming galaxies, especially at the highest radio luminosities, but the difference is much less noticeable than in the previous q$_{\rm FIR}$ distributions presented in Figures \ref{fig:q100} and \ref{fig:q160}. We can then conclude that, even at MIR levels, radio-selected AGN and star-forming galaxies are roughly indistinguishable from each other.
Interestingly enough, both populations show a marked preference to follow the tracks highlighted by the Arp220 SED. This implies that, independent of the origin of radio activity, their hosts are the sites of extremely intense ongoing star-formation processes, like it is the case for Arp220-like galaxies.

%The reason for this result probably resides in the fact that the 24$\mu$m emission is in most  cases powered by AGN accretion. This is also probably the reason why a trend like that of Arp220 (which is known to host a central AGN, e.g. Spoon et al. 2004 and references therein) is preferred to that of M82, a more normal star-forming galaxy. Note that the above finding also implies the presence of a radio-emitting AGN at least in a fraction of the P$_ {1.4 \rm GHz}\simgt 10^{21}$ [W Hz sr$^{-1}$] star-forming galaxy population.

\section{conclusions}
With the aim of investigating the Far-Infrared properties of radio-active AGN, we have cross-matched the COSMOS-VLA catalogue of radio sources with fluxes F$_{1.4 \rm GHz}\ge 60 \mu$Jy with two datasets of galaxies observed at 100$\mu$m and 160$\mu$m by the {\it Herschel}-PEP survey.

In order to minimize biases due to selections at multiple wavelengths, AGN were extracted from the COSMOS-VLA sample only on the basis of their radio luminosity. This was done by following the recent results of McAlpine et al. (2013) for the evolution of the radio luminosity function of star-forming galaxies and AGN up to $z\sim 2.5$.

 To achieve our goal, we have then considered the 1537 radio sources (65\% of the parent sample, independent of radio flux) from the COSMOS-VLA survey which also had a reliable redshift estimate.  Thanks to the depth of the radio survey, our selection criteria ensure that the radio-active AGN population is {\it complete} with respect to radio selection up to redshifts $z\sim 3.5$.

Out of the 1537 objects from the COSMOS-VLA survey with a redshift estimate, 832 were found to have a FIR counterpart, either at 100$\mu$m (F$_{100\mu \rm m}\simgt 4$ mJy)  or at 160$\mu$m (F$_{160\mu \rm m}\simgt 7$ mJy). 175 of these sources were classified as AGN according to our selection criteria. This corresponds to $\sim 36$\% of the parent, radio-selected, AGN sample. 

The redshift distribution of radio-selected AGN with a  FIR counterpart closely follows that of the parent AGN population and shows two marked peaks at $z\sim 0.9$ and $z\sim 2.5$. On the contrary, irrespective of their FIR emission, the redshift distribution of radio-selected star-forming galaxies is found to monotonically decrease with look-back time.

We find that the  probability for a radio-selected AGN to be detected at FIR wavelengths is a strong function of radio luminosity, whereby more luminous sources have the lowest chances of also being FIR emitters. 
However, as there is also a strong dependency on redshift,  such chances are minimal for powerful radio sources in the local universe. For instance,  a typical radio AGN with P$_{1.4 \rm GHz}\simgt 10^{23}$ [W Hz sr$^{-1}$] has $\sim 10-20$\% probability of being detected at FIR wavelengths for $z<1$, while this percentage rises to $\sim 40$\% in the redshift range $z=[1-2]$, and up to $\sim 50-60$\% at the highest redshifts probed by our analysis. We have shown that this is due to a combination of two effects: the first one is the enhancement of FIR activity with look-back time observed in radio-selected AGN of all luminosities. This effect is further boosted by the fact that at earlier epochs radio activity is found to be less efficient at inhibiting the processes which are responsible for FIR emission. 

By analysing the mass distribution of radio-selected AGN with or without a FIR counterpart, we also found that FIR emitters are associated to galaxies of statistically lower stellar masses. Also, there seems to exist a quite narrow range of masses, 
M$_*\sim[10^{10}-10^{11}]$ M$_\odot$, which maximizes the chances for FIR emission to about 50\% or more of the parent sample.  Beyond such range, the probability sharply drops to $\simlt$20\%. However, this is only valid for $z\simlt 2$. At higher redshifts, we do not observe any preferential mass scale.

By comparing the observed FIR, MIR and radio colours with a range of available templates, we can conclude that FIR activity in radio-selected AGN is due to processes connected to star-forming activity within the host galaxy. Such processes produce FIR emission which is indistinguishable from that produced in "pure" star-forming galaxies. An important point to stress is that since star-forming processes also generate radio emission,  the observed radio signal produced by radio-selected AGN which are also active at FIR wavelengths is generated by the superposition of two contributions, one due to AGN accretion and one caused by stellar formation within the host galaxy. 

Note that the above results refer to the $\sim 36$\% of  radio-selected AGN with a FIR counterpart in the PEP survey. However, since  the AGN sample is basically complete in radio selection at all redshifts, and since the fraction of  AGN which have a FIR counterpart is  found to be roughly independent of both redshift and radio flux, we expect that probing FIR fluxes that are fainter than those observed by the PEP survey would merely have the effect of increasing the fraction of FIR-detected sources, while leaving  the general conclusions of this paper unchanged. We have also shown that our results are safe against possible contamination of the AGN sample with star-forming galaxies.

Putting together the various pieces of information we have gathered so far then returns  a very interesting picture on the cosmological evolution of the population of radio-active AGN: at redshifts which coincide with the peak of cosmic star-formation, $z\sim [2-3]$, many of these sources (about 50\% from our sample, but as previously discussed we expect more from deeper FIR surveys) are associated to strong FIR emission due to intense star-forming activity within the host galaxy. These concomitant processes of star-formation and AGN accretion take place in all galaxies with stellar masses M$_*\simgt 10^{10}$ M$_\odot$ without a preferential mass scale, and star-formation is only inhibited in the very few extremely powerful, P$_{1.4 \rm GHz}\simgt 10^{25}$ [W Hz sr$^{-1}$], AGN. This happens not only because cosmic star-formation is maximum at these epochs even in radio-selected AGN, but also because the process is so intense that even important feedback of AGN origin cannot swipe off gas from the galaxy  sites  where star formation takes place. 

As we approach the more local universe, things change. FIR activity in AGN hosts decreases and even a small amount of AGN feedback (like that produced by a P$_{1.4 \rm GHz}\simeq 10^{23}$ [W Hz sr$^{-1}$] source), becomes increasingly more effective at switching off stellar production in the host galaxy. Radio-active AGN are progressively found associated to more and more "red and dead" galaxies of large masses. Concomitant star-formation and AGN activity only remain in a very small fraction ($\simlt$20\% in our sample for $z\simlt 1$) of objects which are hosted by relatively small galaxies and whereby AGN emission and its related feedback are increasingly weaker. Interestingly enough, our results mirror those obtained for AGN selected in the X-ray band (Page et al. 2012). 
This implies that the cosmic history of AGN feedback is a general feature which does not depend on the waveband adopted to select black-hole powered sources.

The above results put a warning to all those criteria which use MIR and/or FIR information to distinguish between AGN-powered sources and  star-forming galaxies. In fact, while these methods hold in the local, $z\simlt 1$, universe, the concomitant presence of AGN accretion and star formation in the majority of the sources selected at the radio wavelengths at higher redshifts implies that a clear distinction between these two populations based on their infrared properties might prove to be much harder than previously thought.\\
\\

\noindent
{\bf Awcknoledgements}
We wish to  thank the anonymous referee for the constructive comments which  helped  improving the paper.


\begin{thebibliography}{}
\bibitem[\protect\citename{alexander}2008]{alexander}
Alexander et al. 2008, ApJ, 135, 1968
\bibitem[\protect\citename{appleton}2004]{app}
Appleton P.N. et al., 2004, ApJS, 154, 147
\bibitem[\protect\citename{Berta}2010]{berta}
Berta S. et al., 2010, A\&A, 518, L30
\bibitem[\protect\citeauthoryear{Bondi1}2003]{Bondi1}
Bondi M. et al. 2003, A\&A, 403, 857
\bibitem[\protect\citeauthoryear{Bondi}2008]{Bondi}
Bondi M., Ciliegi P., Schinnerer E., Smolcic V., Jahke K., Carilli C., Zamorani G., 2008, ApJ, 681, 1135
\bibitem[\protect\citeauthoryear{Bourne}2011]{Bourne}
Bourne N., Dunne L., Ivison R.J., Maddox S.J., Dickinson M., Frayer D.T., 2011, MNRAS, 410, 1155
\bibitem[\protect\citename{Boyle} 2007]{Boyle}
Boyle B.J., Cornwell T.J., Middleberg E., Norris R.P., Appleton P.N., Smail I., 2007, MNRAS, 385, 1143
\bibitem[\protect\citeauthoryear{Carillii}2004]{Carilli}
Carilli C.L., Furlanetto S., Briggs F., Jarvis M., Rawlings S., Falcke H., 2004, New Astronomy Reviews, Elsevier, Vol. 48, Issue 11-12, 1029
\bibitem[\protect\citename{Condon2}1982]{Condon2}
Condon J.J., Condon M.A., Gisler G., Pushell J., 1982, ApJ, 252, 102
\bibitem[\protect\citename{Condon3}1992]{Condon3}
Condon JJ., 1992, ARA\&A, 30, 575
\bibitem[\protect\citename{De Breuck} 2010]{Deb}
De Breuck C. et al., 2010, ApJ, 725, 36
\bibitem[\protect\citename{Del Moro} 2013]{Del}
Del Moro A. et al., 2013, A\&A, 549, A59
\bibitem[\protect\citename{Dickey}1984]{Dickey}
Dickey J.M. \& Salpeter E.E., 1984, ApJ, 284, 461
\bibitem[\protect\citename{Fanaroff} 1974]{Fanaroff}
Fanaroff  B.L., Riley J.M., 1974, MNRAS, 167, 31
\bibitem[\protect\citename{Fine}2011]{Fine}
FIne S. Shanks T., Nikoloudakis N., Sawangwit U., 2011, MNRAS, 418, 2251
\bibitem[\protect\citename{Garn}2009]{Garn}
Garn T., Green D.A., Riley J.M. Alezander P., 2009, MNRAS, 397, 1101
\bibitem[\protect\citeauthoryear{Griffin}2010]{griffin}
Griffin M.J. et al., 2010, A\&A, 518, L3
\bibitem[\protect\citeauthoryear{Gruppioni}2013]{gruppioni}
Gruppioni C. et al. 2013, MNRAS, 432, 23
\bibitem[\protect\citename{Ibar}2010]{Ibar}
Ibar E., Ivison R.J., Best P.N., Coppin K., Pope A., Smail I., Dunlop J.S., 2010, MNRAS, 401, L53
\bibitem[\protect\citeauthoryear{Ilbert}2012]{Ilbert}
Ilbert O., et al., 2013, A\&A, 2013, 556, A55
\bibitem[\protect\citename{Ivison}2010]{Ivison}
Ivison R.J., et al., 2010, A\&A, 518, L31
\bibitem[\protect\citename{Johnston}2007]{Johnston}
Johnston S. et al., 2007,  PASA, 24, 174
\bibitem[\protect\citename{Jonas}2009]{Jonas}
Jonas J.L., 2009, IEEE Proc. 97, 1522
\bibitem[\protect\citename{Leip} 2010]{Leip}
Leipski C., Antonucci R., Ogle P., Whysong D., 2009, 701, 891
\bibitem[\protect\citename{Lilly}1984]{Lillly}
Lilly S.J., Longair M.S., 1984, MNRAS, 211, 833
\bibitem[\protect\citename{Lutz}2011]{Lutz}
Lutz D. et al., 2011, A\&A, 532, 90
\bibitem[\protect\citename{Maglio}1999]{Maglio}
Magliocchetti M.,  Maddox S. J., Wall J. V., Benn C. R., Cotter G. 2000, MNRAS, 318, 1047
\bibitem[\protect\citename{Maglio1}2002]{Maglio1}
Magliocchetti M. et al. (the 2dFGRS Team) 2002, MNRAS, 333, 100
\bibitem[\protect\citename{Maglio2}2004]{Maglio2}
Magliocchetti M. et al. (the 2dFGRS Team) 2004, MNRAS, 350, 1485
\bibitem[\protect\citename{Maglio3}2008]{Maglio3}
Magliocchetti M., Andreani P., Zwaan M.A., 2008, MNRAS, 383, 479
\bibitem[\protect\citename{Mao}2011]{Mao}
Mao M.Y. et al., 2011, ApJ, 731, 79
\bibitem[\protect\citename{Mao}2007]{Mao}
Mao M.Y. et al., 2012, MNRAS, 426, 3334
\bibitem[\protect\citename{Mauch}2007]{Mauch}
Mauch T., Sadler E.M., 2007, MNRAS, 375, 931
\bibitem[\protect\citename{McAlpine}2012]{McAlpine}
McAlpine K., Jarvis M.J., Bonfield D.G., 2013, MNRAS, 436, 1084
\bibitem[\protect\citename{Merloni}2004]{Merloni}
Merloni A., Rudnik G., DI Matteo T., 2004, MNRAS, 354, L37
\bibitem[\protect\citename{Metcalfi}2006]{Metcalf}
Metcalf R.B., Magliocchetti M., 2006, MNRAS, 365, 101
\bibitem[\protect\citename{Morrison}2010]{Morrison}
Morrison G. E., Ower F.N., Dickinson M., Ivison R.J., Ibar E., 2010, ApJS, 188, 178
\bibitem[\protect\citename{Norris} 2011]{Norris}
Norris R.P. et al., 2011, ApJ, 736, 55
\bibitem[\protect\citename{Page} 2012]{page}
Page M.J. et al., 2012, Nature, 485, 213
\bibitem[\protect\citename{Pil} 2010]{pil}
Pilbratt G. et al., 2010, A\&A, 518, L1
\bibitem[\protect\citename{Pog} 2010]{pog}
Poglitsch A. et al., 2010, A\&A, 518, L2
\bibitem[\protect\citename{Ran} 2011]{Ran}
Randall K.E. , Hopkins A.M., Norris R.P., Zinn P.-C., Middleberg E., Mao M.Y., Sharp R.G., 2012, MNRAS, 421, 1644
\bibitem[\protect\citename{Sargent}2010]{Sargent}
Sargent M.T., et al., 2010, ApJSS, 186, 341
\bibitem[\protect\citeauthoryear{Schin}2004]{Schin1}
Schinnerer E., et al., 2004, AJ, 128, 1974
\bibitem[\protect\citeauthoryear{Schin}2007]{Schin2}
Schinnerer E. et al.., 2007, ApJS, 172, 46
\bibitem[\protect\citeauthoryear{Schin}2010]{Schin3}
Schinnerer E. et al.., 2010, ApJS, 188, 384
\bibitem[\protect\citeauthoryear{seymour}2011]{seymour}
Seymour N. et al., 2011, MNRAS, 413, 1777
\bibitem[\protect\citeauthoryear{Simp}2012]{Simp}
Simpson C. et al. 2012, MNRAS, 421, 3060
\bibitem[\protect\citeauthoryear{Smolcic1}2009a]{Smol1}
Smolcic V.et al 2009a, ApJ, 690, 610
\bibitem[\protect\citeauthoryear{Smolcic2}2009b]{Smol2}
Smolcic V.et al 2009b, ApJ, 696, 24
%Spoon H.W.W., Moorwood A.F.M., Lutz D., Tielens A.G.G.M.,  Siebenmorgen R.,  Keane J.V., 2004, A\&A, 414, 873
\bibitem[\protect\citeauthoryear{Urry}1995]{Urry}
Urry M. \& Padovani P., 1995, PASP, 107, 803
\bibitem[\protect\citeauthoryear{White}20012]{White}
White G.J. et al. 2012, MNRAS, 427, 1830

\end{thebibliography}
\end{document}